\documentclass[12pt,reqno]{amsart}
\usepackage{amsmath,amsfonts,amssymb,amscd,amsthm,amsbsy,epsf}
\usepackage[dvips]{graphicx}
\usepackage[usenames]{xcolor}
\usepackage{float}

\numberwithin{equation}{section}

\newtheorem{theorem}{Theorem}

\newtheorem{definition}[theorem]{Definition}

\newcommand\beq[1]{ \begin{equation}\label{#1} }
\newcommand{\eeq}{ \end{equation} }

\newcommand\beqa[1]{ \begin{eqnarray} \label{#1}}
\newcommand{\eeqa}{ \end{eqnarray} }
\newcommand{\beqano}{ \begin{eqnarray*} }
\newcommand{\eeqano}{ \end{eqnarray*} }
\newcommand\equ[1]{{\rm (\ref{#1})}}

\def\G{{\mathcal G}}

\def\integer{{\mathbb Z}}

\def\real{{\mathbb R}}
\def\torus{{\mathbb T}}

\begin{document}

\title[Proper Elements for Space Debris]
{Proper Elements for Space Debris}

\author[A. Celletti]{Alessandra Celletti}
\address{
Department of Mathematics, University of Roma Tor Vergata, Via della Ricerca Scientifica 1,
00133 Roma (Italy)}
\email{celletti@mat.uniroma2.it}

\author[G. Pucacco]{Giuseppe Pucacco}
\address{
Department of Physics, University of Roma Tor Vergata, Via della Ricerca Scientifica 1,
00133 Roma (Italy)}
\email{pucacco@roma2.infn.it}

\author[T. Vartolomei]{Tudor Vartolomei*}
\address{
Department of Mathematics, University of Roma Tor Vergata, Via
della Ricerca Scientifica 1, 00133 Roma (Italy)}
\email{vartolom@mat.uniroma2.it}

\baselineskip=18pt              

\begin{abstract}
Proper elements are quasi-invariants of a Hamiltonian system, obtained through a
normalization procedure. Proper elements have been successfully used to identify
\sl families \rm of asteroids, sharing the same dynamical properties. We show that proper
elements can also be used within space debris dynamics to identify \sl groups \rm of fragments
associated to the same break-up event. The proposed method allows to reconstruct the evolutionary
history and possibly to associate the fragments to a parent body.

The procedure relies on different steps: (i) the development of a model for an approximate, though accurate,
description of the dynamics of the space debris; (ii) the construction of a normalization
procedure to determine the proper elements;
(iii) the production of fragments through a simulated break-up event.

We consider a model that includes the Keplerian part, an
approximation of the geopotential, and  the gravitational influence of
Sun and Moon. We also evaluate the contribution of Solar radiation pressure
and the effect of noise on the orbital elements.
We implement a Lie series normalization procedure to
compute the proper elements associated to semi-major axis,
eccentricity and inclination. Based upon a wide range of samples,
we conclude that the distribution of the proper elements in
simulated break-up events (either collisions and explosions) shows
an impressive connection with the dynamics observed immediately
after the catastrophic event.
The results are corroborated by a statistical data analysis based on the
check of the Kolmogorov-Smirnov test and the computation of the Pearson
correlation coefficient.
\end{abstract}

\maketitle

\vglue.1cm

\noindent \bf Keywords. \rm Proper elements, Normal form, Space
debris, Geopotential, Sun and Moon attractions, Solar radiation pressure,
Noise, Statistical data analysis.

\vglue.1cm

\section{Introduction}\label{sec:intro}

Over the centuries perturbation theory has been used in different contexts of
Celestial Mechanics, from the computation of the ephemerides of the Moon to
the determination of the orbits of artificial satellites.

In the domain of nearly-integrable Hamiltonian systems,
perturbation theory consists in a normalization procedure that
implements a canonical change of coordinates so that the
transformed system becomes integrable up to a remainder function.
The integrable part of the transformed Hamiltonian admits
integrals of motion, which are indeed \sl quasi-integrals \rm for
the Hamiltonian system that includes the remainder. The
normalization procedure is \sl constructive \rm in the sense that
it allows us to determine the explicit expression of the new integrable
part and, hence, of the quasi-integrals.

This procedure has been successfully used to compute the so-called
\sl proper elements, \rm which are quasi-invariants of the dynamics,
staying nearly constant over very long times. Proper elements found a
striking application in the context of the grouping of asteroids
to form \sl families. \rm We refer to \cite{K2016IAU} for a thorough review
of the history of asteroid family identification.
The idea at the basis of such computation is that
objects with nearby proper elements might have been physically close
in the past. One can even conjecture that such asteroids might be
fragments of an ancestor parent body. Being obtained through the averaging method over short-period angles followed by a normal form procedure that averages out the long-period perturbations, proper elements retain the
essential features of the original family formation, which can be lost when using the osculating elements at the present time.

Proper elements were used in 1918 by Hirayama
(\cite{Hirayama}), who noticed many asteroids with similar
semi-major axes forming ring-shaped clusters in projections on the planes
of non-singular equinoctial orbital elements; the radius of the
ring represents a proper element. Later, Brouwer (\cite{Brouwer})
computed proper elements using an improved theory of planetary
motion, while Williams (\cite{Williams}) developed a semianalytic
theory of asteroid secular perturbations. Using Yuasa theory
(\cite{Yuasa}), Kozai identified asteroid families formed by
high-inclination asteroids (\cite{Kozai}, \cite{LM}, \cite{Novacovic2011}).
Proper elements for celestial bodies in resonance, e.g. Trojan
asteroids, were considered by Schubart (\cite{Schubart}, see also
\cite{Morbidelli1993}). A number
of relevant works that included an extension of Yuasa theory were
developed by Kne{\v z}evi{\'c} and Milani (see, e.g.,
\cite{MK1990}, \cite{MK1994}, see also \cite{Lemaitre1992}). The
latter authors widely discussed also an alternative method to
compute the proper elements, based on a so-called \sl synthetic
\rm theory, which uses a numerical integration, a digital
filtering of the short-period terms and a Fourier analysis
(\cite{KnezevichLemaitreMilani2002}, \cite{KnezevichMilani2000}, \cite{KnezevichMilani2003}, \cite{KnezevichMilani2019}).

\vskip.1in

Inspired by the results of the computation of proper elements for asteroids,
in this work we aim at computing and testing proper elements for the space debris problem.
The increasing number of space debris surrounding the Earth has become
a serious threat for the safeguard of operative satellites and for
space missions. Many efforts are concentrated to track and
prevent catastrophic events involving space debris, which
are normally hardly observable and have high velocities.
Here, we develop a method to classify the
space debris into families, such that one can recover useful
information about their evolution.

\vskip.1in

The procedure can be split into the following main steps:
(i) the development of a model that, in a given region, provides a good approximation
of the dynamics of the space debris; (ii) using Lie series,
the explicit construction of an iterative normalization procedure to determine the proper elements;
(iii) the production of fragments through a simulated break-up event and their analysis
by means of a comparison between initial osculating elements, mean elements after a given period of time, and proper elements computed using the mean elements.

We provide below some details of the above three steps.

\subsection{The model}

\begin{figure}[h]
\centering
\includegraphics[width=1\columnwidth]{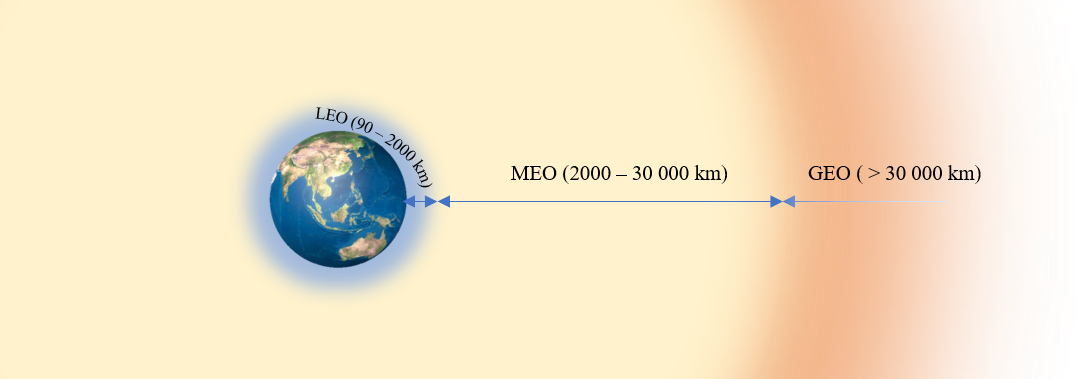}
\caption{LEO: 90-2000 km, MEO: 2000-30000 km, GEO: $\geq$ 30000 km}
\label{fig:regions}
\end{figure}

To study the dynamics around the Earth, it is convenient to split
the region surrounding our planet into three main parts:
Low-Earth-Orbits (LEO) between 90 and 2000 km of altitude,
Medium-Earth-Orbits (MEO) between $2000$ and $30000$ km of altitude,
Geosynchronous Earth's Orbits (GEO) above $30000$ km of altitude
(see Figure~\ref{fig:regions}).

The motion in LEO-MEO-GEO is mainly governed by the gravitational field of the Earth,
that must include also the fact that the shape of our planet is non spherical;
in LEO it is important to consider the dissipative effect due to the atmospheric drag,
while in MEO and GEO the gravitational influence of Moon and Sun,
as well as the Solar radiation pressure (hereafter, SRP), play a very important role
(see, e.g., \cite{CPL2015}, \cite{CGLEO}, \cite{cellettietal2020}, \cite{GC},
\cite{Lho2016}, \cite{SARV}, \cite{SRT}).
We will limit our study to MEO and GEO, in which the dynamics is
governed by a conservative model.
The computation of quasi-invariants in GEO has been also approached in
\cite{cellettietal2017rev}, \cite{gachetetal2017}.

In this work, we introduce a Hamiltonian function describing
an approximation of the contribution of the sum of the Keplerian part,
an expansion of the geopotential, and the gravitational influence of
Moon and Sun. The resulting Hamiltonian depends upon the orbital
elements of the debris, Moon, Sun and on the sidereal time accounting
for the rotation of the Earth. Hence, the Hamiltonian depends upon
a set of angular variables, that can be ordered hierarchically as
fast, semi-fast and slow variables, since their rates change over days, months, years.
We also investigate the effect of Solar radiation pressure, providing
details on a specific sample.

The model is described in Section~\ref{sec:Model}.

\subsection{Normal forms}

The normalization procedure is based upon the following strategy.
We first average the model Hamiltonian over the fast and semi-fast angles;
as a consequence of such averaging, the semi-major axis is constant and
becomes the first proper element.
As a result, we obtain a two degrees of freedom Hamiltonian, which depends
on time, due to the variation of the longitude of the ascending node of the Moon.
After the introduction of the extended Hamiltonian, we make an expansion around
reference values of the action variables. Next, we proceed to implement a
Lie series normalization to first order that allows us to determine
two more proper elements, corresponding to eccentricity and inclination.

The normalization procedure is presented in Section~\ref{sec:Normalization}.

\subsection{Analysis of fragments' clusters}

To analyze groups of fragments originated from the same parent body, we
use a simulator of break-up events based upon the program developed in
\cite{simulator}. We generate fragments from explosions or collisions
of two satellites and we consider fragments with size greater than 12 cm.
We consider break-up events occurring in three regions with different semi-major axis $a$:
moderate altitude orbits ($a=15600$ km and $a=20600$ km), and
medium altitude orbits ($a=33600$ km).

Our results consist in a comparison of the three sets of data
providing semi-major axis, eccentricity and inclination, obtained
as follows: we consider the elements just after break-up,
after a propagation over 150 years, and we compute the proper elements
based on the data propagated after 150 years.
The results are presented in Section~\ref{sec:ProperElements},
which shows also some results obtained propagating the fragments
to times less than 150 years. The distribution of the proper elements is
analyzed by making the computations at different propagation times
and this is supported by statistical data analysis by drawing histograms, by performing the Kolmogorov-Smirnov test and by computing the Pearson correlation coefficient.

This analysis is performed in non-resonant regions, but also
resonant regions might be interesting as well. To this end, we
introduce the definition of \sl tesseral resonance, \rm which
corresponds to a commensurability involving the rate of variation of
the mean anomaly of the debris and the rotation of the Earth. In
Section~\ref{sec:ProperElementsTes21}, we extend the computation
to the study of objects in the vicinity of resonant regions, precisely the 1:1 and
2:1 tesseral resonances. Finally, we complement the results
by providing a sample in which we add the effect of noise to
the initial mean elements.

\vskip.1in

From the results obtained in the non-resonant and resonant regions,
we can draw the following conclusions: in several cases, the proper
elements allow us to reconstruct a cluster structure, very similar to that
which is formed just after the break-up. In particular, in the non-resonant
case, the proper elements are an efficient tool to reconstruct the initial distribution.
It is worth mentioning that some cases might be  affected by lunisolar resonances
(\cite{sB01}, \cite{cellettigalespucacco}, \cite{CellettiGales2016}, \cite{EH},
\cite{HughesI}) and therefore the cluster configuration can show anomalies
in the distribution of the fragments.

\vskip.1in

The method presented in this paper represents an important tool for
the classification of space debris into families; it is clear that such result has a major
impact in many directions of paramount importance in space debris
dynamics, most notably the identification of the space debris parent body.

Some further conclusions and perspectives of this work are given
in Section~\ref{sec:conclusions}.

\section{The model} \label{sec:Model}

In this Section we introduce a model for the description of the
dynamics of space debris, that takes into account four main
contributions: the gravitational potential of the Earth,
the attraction of the Moon, the attraction of the Sun, and the Solar radiation pressure. The model
has been described in full detail in \cite{cellettietal2017rev},
\cite{CellettiGales2014}, starting from the Cartesian equations
of motion and using the expansion in orbital elements.\\

With reference to \cite{cellettietal2017rev}, \cite{CellettiGales2014}, we consider a Hamiltonian
function composed by the following parts:
$$
\mathcal{H} = \mathcal{H}_{Kep} + \mathcal{H}_E + \mathcal{H}_M+ \mathcal{H}_S + \mathcal{H}_{SRP} \ ,
$$
where $\mathcal{H}_{Kep}$ denotes the Keplerian part due to the
interaction with a spherical Earth, $\mathcal{H}_E$
is the contribution due to the non-spherical Earth,
$\mathcal{H}_M$ and $\mathcal{H}_S$ denote,
respectively, the Hamiltonian parts describing the attractions of
Moon and Sun, $\mathcal{H}_{SRP}$ denotes the effect
of Solar radiation pressure.

The Keplerian part takes a simple form, while
the Hamiltonian functions $\mathcal{H}_E$, $\mathcal{H}_M$, $\mathcal{H}_S$, $\mathcal{H}_{SRP}$ have more
complex expressions.

\subsection{Keplerian Hamiltonian} \label{sec:kepham}

The Keplerian Hamiltonian can be written in the form
$$
\mathcal{H}_{Kep}(a) = -\frac{\mu _{E}}{2 a}\ ,
$$
where $a$ denotes the semi-major axis, $\mu_E= \mathcal{G}m_E$ with $\mathcal{G}$ the gravitational constant and $m_E$ the mass
of the Earth.

\subsection{Gravitational field of the Earth} \label{sec:GravField}

Following \cite{cellettietal2017rev}, \cite{CellettiGales2014}, \cite{kaula1966}, the Hamiltonian part
corresponding to the geopotential perturbation - assuming that the Earth is not spherical -
can be written as an expansion in the orbital elements $(a,e,i,M,\omega,\Omega)$
of the space object and depending on the sidereal time $\theta$, which takes into account the
rotation of the Earth.

\vskip.1in

In a quasi-inertial reference frame (see Appendix A for more details), the Hamiltonian function can be written as
\begin{multline}\label{eq:EarthExp}
\mathcal{H}_{\text{E}}(a,e,i,M,\omega,\Omega,\theta) = -\frac{\mu_{E}}{a}\sum\limits^{\infty}_{n=2}\sum\limits^{n}_{m=0}(\frac{R_E}{a})^n\sum\limits^{n}_{p=0}F_{nmp}(i)\sum\limits^{\infty}_{q=-\infty}G_{npq}(e)S_{nmpq}(M,\omega,\Omega,\theta),
\end{multline}
where $R_E$ is the Earth's radius and
the functions $F_{nmp}(i)$ and $G_{nmp}(e)$ are given by
\beqa{eq:EarthExpAux}
F_{nmp}(i) &=& \sum\limits_{w=0}^{min(p,K)}\frac{(2n-2w)!}{w!(n-w)!(n-m-2w)!2^{2n-2w}}\sin^{n-m-2w}(i)
\sum\limits^{m}_{s=0}{m\choose s}\cos^s(i) \nonumber\\
&& \sum\limits^{n}_{c=0}{n-m-2w+s\choose c}{m-s\choose p-w-c}(-1)^{c-K}\ ,\nonumber\\
G_{nmp}(e) &=&  (-1)^{|q|}(1+\beta^2)^n\beta^{|q|}\sum\limits_{k=0}^{\infty}P_{npqk}Q_{npqk}\beta^{2k}
\eeqa
with $K=[(n-m)/2]$ and $\beta$, $P_{npqk}$, $Q_{npqk}$ being functions of the eccentricity:
\beqano
\beta &=& \frac{e}{1+\sqrt{1-e^2}}\\
P_{npqk} &=&   \sum\limits_{r=0}^h {2p'-2n\choose h-r}\frac{(-1)^r}{r!}\left(\frac{(n-2p'+2q')e}{2\beta}\right)^r , \begin{cases} h=k+q', q'>0 \\  h=k, q'<0\end{cases} \\
Q_{npqk} &=& \sum\limits_{r=0}^h {-2p'\choose h-r}\frac{1}{r!}\left(\frac{(n-2p'+2q')e}{2\beta}\right)^r , \begin{cases} h=k, q'>0 \\  h=k-q', q'<0\ ,\end{cases}
\eeqano
while $p'=p$ and $q'=q$ when $p\leq n/2$, $p'=n-p$ and $q'=-q$ when $p>n/2$.\\

The quantity $S_{nmpq}$ in (\ref{eq:EarthExp}) is given by
\begin{equation*}
S_{nmpq} =
\begin{cases}
-J_{nm}\cos(\Psi_{nmpq}),\qquad \mod(n-m,2) = 0 \\
-J_{nm}\sin(\Psi_{nmpq}),\qquad \mod(n-m,2) = 1\ ,
\end{cases}
\end{equation*}
where
\begin{equation}\label{eq:EarthAnglesComb}
\Psi_{nmpq}=(n-2p)\omega + (n-2p+q)M + m(\Omega-\theta)-m\lambda_{nm}
\end{equation}
for some constants $\lambda_{nm}$, while $J_{nm}$ are related to the
spherical harmonic coefficients of the geopotential (\cite{kaula1962}).
According to a standard notation, we define $J_n\equiv J_{n0}$. We remark
that among the coefficients $J_{nm}$ of the Earth the biggest one is
$J_2=-1.082\cdot 10^{-3}$; the values of the first few coefficients $J_{nm}$ and $\lambda_{nm}$ are
listed in Table~\ref{table:J}.

\vskip.2in

\begin{table}[h]
\begin{tabular}{|c|c|c|c|}
  \hline
  $n$ & $m$ & $J_{nm}$ & $\lambda_{nm}$ \\
  \hline
  2 & 0 & $-1082.2626$& 0 \\
  2 & 1 & 0.001807& $-81_{\cdot}^{\circ}5116$ \\
  2 & 2 & 1.81559& $75_{\cdot}^{\circ}0715$ \\
  3 & 0 & $-2.53241$& 0 \\
3 & 1 & 2.20947& $186_{\cdot}^{\circ}9692$ \\
3 & 2 & 0.37445& $72_{\cdot}^{\circ}8111$ \\
3 & 3 & 0.22139& $80_{\cdot}^{\circ}9928$ \\
  \hline
 \end{tabular}
 \vskip.1in
 \caption{Values of $J_{nm}$ (in units of $10^{-6}$) and $\lambda_{nm}$ from \cite{EGM2008}.}\label{table:J}
\end{table}

\vskip.2in

Instead of the full expansion (\ref{eq:EarthExp}), we will consider only the secular part by averaging
over the fast angles $M$ and $\theta$; this amounts to choose in (\ref{eq:EarthAnglesComb}) the terms
with indexes $m=0$ and $n-2p+q=0$. The explicit expansions of $\mathcal{H}_E$ up to order $n=2$ or $n=3$, and including
only the $J_2$ term, take the following form:
$$
\mathcal{H}_E^{J_2}=\mu _{\text{E}} R_{\text{E}}^2 J_2 \frac{1+3 \cos (2 i)}{8 a^3 \left(1-e^2\right)^{3/2}},
$$
while including both $J_2$ and $J_3$ terms, one obtains
\beqa{eq:HamJ3Averaged}
\mathcal{H}_E^{J_3}&=&{{\mu _{\text{E}} R_{\text{E}}^2}\over a^3}\ J_2\ ({3\over 4}\sin^2 i-{1\over 2}){1\over {(1-e^2)^{3\over 2}}}\nonumber\\
&+&{{\mu _{\text{E}} R_{\text{E}}^3}\over a^4}\ J_3\ ({{15}\over 8}\sin^3 i-{3\over 2}\sin i)e\sin\omega\ {1\over {(1-e^2)^{5\over 2}}}\ .
\eeqa
We give now the following definition of \sl tesseral resonance, \rm which plays an important role both in theoretical investigations
and practical applications to Earth's satellites.

\vskip.1in

\begin{definition}\label{def:tesseral}
A tesseral resonance of order j:l with $j, l \in \mathbb{Z} \backslash \{0\}$ occurs whenever
the following relation is satisfied
$$
l\dot{M}-j\dot{\theta}+j\dot{\Omega}+l\dot{\omega}=0\ .
$$
\end{definition}

\subsection{Moon's perturbation}  \label{sec:MoonAttr}

The perturbation of the space object due to the Moon's attraction can be written as an expansion in the
orbital elements of the Moon and the object, using the following formula (see \cite{cellettietal2017}, \cite{kaula1962}):
\begin{multline}\label{eq:MoonExp}
\mathcal{H}_M = -\G m_M \sum\limits_{l\geq 2}\sum\limits_{m=0}^{l}\sum\limits_{p=0}^{l}\sum\limits_{s=0}^{l}\sum\limits_{q=0}^{l}
\sum\limits_{j=-\infty}^{\infty}\sum\limits_{r=-\infty}^{\infty} (-1)^{m+s} (-1)^{[m/2]}\frac{\epsilon_m \epsilon_s}{2a_M}
\frac{(l-s)!}{(l+m)!}\left(\frac{a}{a_M}\right)^l F_{lmp}(i) \\
F_{lsq}(i_M) H_{lpj}(e) G_{lqr}(e_M) \{(-1)^{t(m+s-1)+1}U_l^{m,-s}
\cos(\phi_{lmpj}+\phi'_{lsqr}-y_s \pi)+ \\ (-1)^{t(m+s)}U_l^{m,-s}\cos(\phi_{lmpj}-\phi'_{lsqr}-y_s \pi)\}\ ,
\end{multline}
where $y_s = 0$, if $s$ mod 2=0, $y_s = \frac{1}{2}$, if $s$ mod 2= 1, $t = (l-1)$ mod 2, and
$$
\epsilon_m =
\begin{cases}
1,& m=0\\
2,& m\in \mathbb{Z}\backslash \{0\}
\end{cases}
$$
$$
\phi_{lmpj} = (l-2p)\omega + (l-2p+j)M+m\Omega
$$
$$
\phi'_{lsqr} = (l-2q)\omega_M + (l-2q+r)M_M+s(\Omega_M-\frac{\pi}{2})\ .
$$

The functions $F_{lmp}(i)$, $F_{lsq}(i_M)$ and $G_{lqr}(e_M)$ have been introduced in \equ{eq:EarthExpAux},
$H_{lpj}(e)$ are the Hansen coefficients and the function $U_l^{m,s}$ has the following form

\begin{equation*}
U_l^{m,s} = \sum\limits_{r=\max(0,-(m+s))}^{\min(l-s,l-m)}(-1)^{l-m-r}{l+m\choose m+s+r}{l-m\choose r}\cos^{m+s+2r}
({\epsilon\over 2})\sin^{-m-s+2(l-r)}({\epsilon\over 2}),
\end{equation*}
where $\epsilon = 23^o26'21.406''$ is the Earth's obliquity.

We specify since now that later we will limit ourselves to consider \equ{eq:MoonExp} expanded to $l=2$.

\subsection{Sun's perturbation and Solar radiation pressure} \label{sec:SunAttr}
As regards the Sun, we proceed with an expansion similar to that of the Moon,
now involving the orbital elements of the Sun and the space object.
Precisely, we obtain that $\mathcal{H}_S$ is given by (see \cite{cellettietal2017},\cite{kaula1962})
\begin{multline}\label{eq:SunExp}
\mathcal{H}_S = -\G m_S \sum\limits_{l\geq 2}\sum\limits_{m=0}^{l}\sum\limits_{p=0}^{l}\sum\limits_{h=0}^{l}
\sum\limits_{q=-\infty}^{\infty}\sum\limits_{j=-\infty}^{\infty}\frac{a^l}{a_S^{l+1}}\epsilon_m\frac{(l-m)!}{(l+m)!}\\
F_{lmp}(i) F_{lmh}(i_S) H_{lpq}(e) G_{lhj}(e_S) \cos(\phi_{lmphqj}),
\end{multline}
where
$$
\phi_{lmphqj} = (l-2p)\omega + (l-2p+q)M - (l-2h)\omega_S - (l-2h+j)M_S + m(\Omega -\Omega_S).
$$
As for the Moon, later we will limit ourselves to consider \equ{eq:SunExp} expanded to $l=2$.

The contribution to the Hamiltonian due to Solar radiation
pressure is given by:
\begin{multline}\label{eq:SRP}
\mathcal{H}_{SRP} = C_r P_r \frac{A}{m} a_S^2
\sum\limits_{l=1}^1\sum\limits_{s=0}^{l}\sum\limits_{p=0}^{l}\sum\limits_{h=0}^{l}
\sum\limits_{q=-\infty}^{\infty}\sum\limits_{j=-\infty}^{\infty}\frac{a^l}{a_S^{l+1}}\epsilon_s\frac{(l-s)!}{(l+s)!}\\
F_{lsp}(i) F_{lsh}(i_S) H_{lpq}(e) G_{lhj}(e_S)
\cos(\phi_{lsphqj})\ ,
\end{multline}
where $A/m$ is the area-to-mass ratio of the object, $C_r$ is the reflectivity coefficient, and $P_r$ is the radiation pressure for an object located at $1 AU$.

\subsection{Validation of the model}  \label{sec:ModelValidation}
We conclude this Section with a validation of the Hamiltonian model in two different cases: a
non-resonant motion and a motion close to a 2:1 tesseral resonance.
For each case, we perform two numerical integrations: the first one using
the Cartesian equations of motion (described by the equations in Appendix A),
the second one using Hamilton's equations for the Hamiltonian function
in which the contribution of the Earth is limited to the $J_2$ term and to
the resonant one denoted by $\mathcal{H}_{E}^{2:1}$:
\begin{equation}\label{eq:AveragedHamiltonian}
\mathcal{H} =  \mathcal{H}_{E}^{J_2} + \mathcal{H}_{E}^{2:1} + \mathcal{H}_{S} + \mathcal{H}_{M}\ .
\end{equation}
To define the term $\mathcal{H}_{E}^{2:1}$ we proceed as follows.
If one studies a region where the evolution of the space object is in resonance with the rotation of the Earth,
which means a commensurability between the angles $M$ and $\theta$ as in Definition~\ref{def:tesseral},
we can select in (\ref{eq:EarthExp}) only the terms that correspond to that specific resonance.
For example, for the 2:1 tesseral resonance, we retain only the terms with $m\neq 0$ and $2(n-2p+q)=m$,
thus obtaining the following expression:
\beqano
\mathcal{H}_{E}^{2:1}&=&\frac{\mu _{\text{E}} R_E^2}{a^3}\
 \left(\frac{9}{8} e J_{22}
\left(2-2 \cos ^2(i)\right) \cos \left(M+2 (\Omega -\theta )-2 \lambda _{22}\right)\right)\nonumber\\
&-&\frac{\mu _E R_E^2}{a^3} \left(\frac{3}{8} e J_{22} \left(\cos ^2(i)+2 \cos (i)+1\right)
\cos \left(M+2 \omega+2 (\Omega -\theta )-2 \lambda _{22} \right)\right)\ .
\eeqano

\vskip.1in

In \equ{eq:AveragedHamiltonian}, we assume that the Hamiltonian
function for the Moon (see (\ref{eq:MoonExp})) is expanded up to
order $l=2$ and we take the orbital elements as in Table~\ref{tab:SunMoon}.
Similarly, the Hamiltonian function for the Sun (see (\ref{eq:SunExp})) is
expanded up to order $l=2$ and we take the orbital elements as
in Table~\ref{tab:SunMoon}.

\begin{figure}[h]
\centering
\includegraphics[width=1\columnwidth]{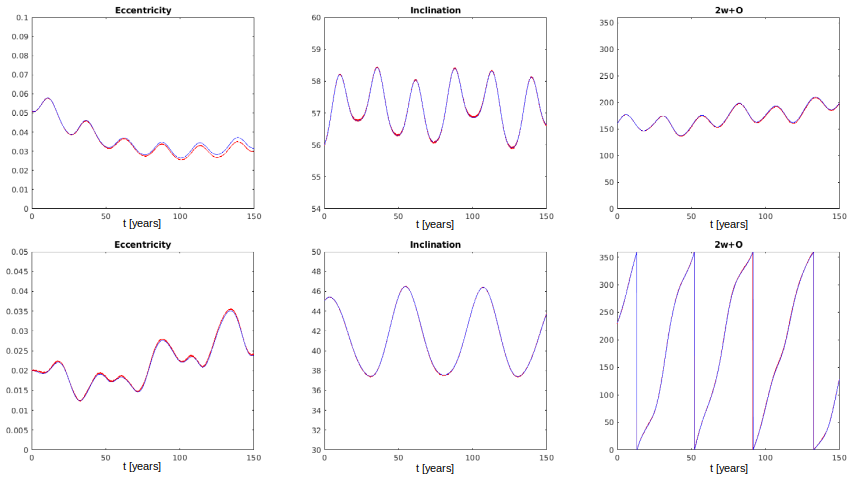}
\caption{Hamilton's integration (red lines) vs Cartesian
Integration (blue lines). Top: $a=26520$ km, $e=0.05122$, $i=56^o$,
$\omega=190^o$, $\Omega=140^o$. Bottom: $a=36700$ km, $e=0.02$,
$i=45^o$, $\omega=100^o$, $\Omega=30^o$.
Left: eccentricity. Middle: inclination. Right: resonant angle $2\omega+\Omega$.} \label{fig:validation1}
\end{figure}

Figure~\ref{fig:validation1} shows the evolution over 150 years
of eccentricity, inclination and resonant angle for two different orbits: the upper one in
a 2:1 resonance and the lower one in a non-resonant region. In each case we plot two
solutions, one obtained integrating the Cartesian equations and
the other one obtained integrating
Hamilton's equations associated to \equ{eq:AveragedHamiltonian}.
The results confirm that the Hamiltonian (\ref{eq:AveragedHamiltonian}) is already accurate enough.
This motivates our choice to use a truncated expansion to order $l=2$ within the
normalization procedure described in
Section~\ref{sec:Normalization} and implemented in
Section~\ref{sec:ProperElements}.


\section{Normalization algorithm} \label{sec:Normalization}

The normalization algorithm is an iterative procedure that
transforms a Hamiltonian function, using canonical coordinates
changes, so that the transformed Hamiltonian
takes a prescribed form, for example that it is
integrable up to a remainder term. The normalization algorithm can be iterated
with the goal to reduce the size of the
remainder term, although we must be aware that the procedure is not converging
in general (\cite{Poin}) and that the computational
complexity increases as the number of normalization steps becomes higher.

We adopt a normalization algorithm based on the use of Lie series
(see,  e.g., \cite{christos2012book});
although it is a standard procedure, we briefly recall the normal
form algorithm, since it is at the basis of the computation of the
proper elements.

\vskip.1in

Let $\mathcal{H}=\mathcal{H}({\underline I},{\underline \varphi})$
be a Hamiltonian function defined in terms of action-angle
variables $({\underline I},{\underline \varphi})\in
B\times\torus^n$, where $B\subset\real^n$ is an open set and $n$
denotes the number of degrees of freedom. We write the Hamiltonian
as
\begin{equation}\label{eq:HamGenInit}
\mathcal{H}({\underline I},{\underline \varphi}) =
\mathcal{H}_0({\underline I}) + \varepsilon
\mathcal{H}_1({\underline I},{\underline \varphi})\ ,
\end{equation}
where $\mathcal{H}_0({\underline I}) $ represents the integrable
part, $\mathcal{H}_1({\underline I},{\underline \varphi})$ is
the perturbing term and $\varepsilon$ represents a small parameter.
We assume that $\mathcal{H}_1$ is the sum of
products between functions depending on actions and cosines of
different combinations of angles; hence, $\mathcal{H}_1$ can be expanded in Fourier series as
\beq{H1}
\mathcal{H}_1({\underline I},{\underline
\varphi})=\sum_{{\underline k}\in \mathcal{K}} b_{\underline k}({\underline I})\exp(\mathit{i} {\underline k}\cdot {\underline \varphi})\ ,
\eeq
where $\mathcal{K}\subseteq\mathbb{Z}^n$ and $b_{\underline k}$ denote functions with real coefficients.

We look for a canonical transformation with generating function
$\chi$ that allows us to perform the change of variables from
$({\underline I},{\underline \varphi})$ to $({\underline
I}',{\underline \varphi}')$ defined through the expressions
\beq{transf}
 {\underline I}=
S_{\chi}^{\varepsilon}{\underline I}'\ ,\qquad {\underline
\varphi} = S_{\chi}^{\varepsilon}{\underline \varphi}'\ ,
\eeq
where the operator $S_{\chi}^{\varepsilon}\mathcal{F}$ is defined by
$$
S_{\chi}^{\varepsilon}\mathcal{F} := \mathcal{F} +
\sum_{i=1}^{\infty}
\frac{\varepsilon^i}{i!}\{\{\dots\{\mathcal{F},\chi\},
\dots\},\chi\}\ ,
$$
and $\{\cdot,\cdot\}$ is the Poisson bracket operator,
such that $\{\mathcal{F},\chi\}=\sum_{j=1}^n
{{\partial \mathcal{F}}\over {\partial \varphi_j}}{{\partial \chi}\over {\partial I_j}}
-{{\partial \mathcal{F}}\over {\partial I_j}}{{\partial \chi}\over {\partial \varphi_j}} $.

The function $S_{\chi}^{\varepsilon}$ must be chosen so that the
transformed Hamiltonian $ \mathcal{H}^{(1)} =
S_{\chi}^{\varepsilon}\mathcal{H}$ takes the following expression:
\begin{equation}\label{eq:HamGenFirstTrans}
\mathcal{H}^{(1)}({\underline I}',{\underline \varphi}') =
\mathcal{H}_0({\underline I}') + \varepsilon
\overline{\mathcal{H}}_1({\underline I}') + \varepsilon^2
\mathcal{H}_2({\underline I}',{\underline \varphi}')\ ,
\end{equation}
where $\mathcal{H}_0+\varepsilon \overline{\mathcal{H}}_1$ is the
new integrable Hamiltonian (depending just on the new actions) and
$\mathcal{H}_2$ is the remainder term (the overbar denotes the average
with respect to the angles).

Inserting the transformation \equ{transf} into
\equ{eq:HamGenInit}, and expanding in Taylor series in the
parameter $\varepsilon$, one obtains that the transformed
Hamiltonian is given by
\beqa{newH}
\mathcal{H}^{(1)}({\underline I}',{\underline \varphi}') &=& \mathcal{H}_0({\underline I}') +
\varepsilon \mathcal{H}_1({\underline I}',{\underline
\varphi}')+\varepsilon\{ \mathcal{H}_0({\underline
I}'),\chi({\underline I}',{\underline \varphi}')\} +
\varepsilon^2\{
\mathcal{H}_1({\underline I}',{\underline \varphi}'),\chi({\underline I}',{\underline \varphi}')\}\nonumber\\
&+&\frac{\varepsilon^2}{2}\{\{\mathcal{H}_0({\underline
I}'),\chi({\underline I}',{\underline
\varphi}')\},\chi({\underline I}',{\underline \varphi}')\} +
\dots
\eeqa
To obtain a Hamiltonian function of the form
(\ref{eq:HamGenFirstTrans}), we must impose that the function in \equ{newH}, that
contains only terms of first order in $\varepsilon$, does not depend on the angles. This
allows us to determine the generating function $\chi$ as the solution of the following homological equation:
\begin{equation}\label{eq:HomologicGen}
\mathcal{H}_1({\underline I}',{\underline \varphi}')+\{ \mathcal{H}_0({\underline I}'),
\chi({\underline I}',{\underline \varphi}')\} =\overline{\mathcal{H}}_1({\underline I}')\ .
\end{equation}

Taking into account the expression \equ{H1} for $\mathcal{H}_1$, we look for a generating function of the form
\begin{equation}
\chi({\underline I}',{\underline \varphi}')=\sum_{\underline{k}\in\mathbb{Z}^n\backslash\{\underline{0}\}}
c_{\underline{k}}({\underline I}')\exp(\mathit{i} \underline{k}\cdot{\underline \varphi}')\ ,
\end{equation}
where the coefficients $c_{\underline{k}}$ will be determined through \equ{eq:HomologicGen}.
In fact, denoting by $\underline{\omega}_0= {{\partial \mathcal{H}_0}\over {\partial{\underline I}'}}$, we obtain
\begin{equation}\label{eq:HomologicalEquation}
\{\mathcal{H}_0({\underline I}'),\chi({\underline \varphi}',{\underline I}')\} =
-\mathit{i}\sum_{\underline{k}\in\mathbb{Z}^n\backslash\{\underline{0}\}}
c_{\underline{k}}({\underline I}')\ \underline{k}\cdot
{\underline\omega}_0\ \exp(\mathit{i} {\underline k}\cdot {\underline
\varphi}')\ .
\end{equation}
Then, equation (\ref{eq:HomologicGen}) is satisfied provided the coefficients $c_{\underline{k}}$ are defined as
\begin{equation}\label{eq:CoefChi}
c_{\underline{k}}({\underline I}')=-\mathit{i}\frac{b_{\underline{k}}({\underline I}')}{\underline{k}\cdot {\underline\omega}_0}\ ,\qquad
{\underline{k}\not=\underline{0}}\ .
\end{equation}
Hence, the generating function takes the form
\beq{chi}
\chi({\underline I}',{\underline
\varphi}')=-\sum_{\underline{k}\in\mathbb{Z}^n\backslash\{\underline{0}\}}
\mathit{i}\frac{b_{\underline{k}}({\underline I}')}{\underline{k}\cdot {\underline\omega}_0}\ \exp(\mathit{i} \underline{k}\cdot{\underline \varphi}')\ .
\eeq
As a consequence, the new Hamiltonian takes the form
(\ref{eq:HamGenFirstTrans}). If one discards the terms of
order $\varepsilon^2$, the normal form is integrable up to orders of $\varepsilon^2$.

If, instead, we keep the terms of order $\varepsilon^2$, we can iterate
the procedure to higher orders to improve the accuracy of the
Hamiltonian normal form. In this case, the new
integrable part is given by $\mathcal{H}_0({\underline I}')+\varepsilon \overline{\mathcal{H}}_1({\underline I}')$ and the perturbation is the
reminder $\varepsilon^2 \mathcal{H}_2({\underline I}',{\underline \varphi}')$.
The algorithm will provide a new generating function that can be constructed
explicitly, using a procedure similar to that leading to \equ{chi}.


\section{Proper elements in a non-resonant region}\label{sec:ProperElements}

In this Section, we apply the normalization algorithm described in
Section~\ref{sec:Normalization} and we compute the proper elements
for several samples of space debris in a non-resonant region of
the phase space; we remind that the proper elements are the
quasi-integrals of the Hamiltonian function describing the
dynamics of the space debris. More precisely, we will compute
exact integrals of the non-resonant normal form  (namely, of the integrable part)
which, when expressed in terms of the original elements, are almost conserved quantities up to order
$\varepsilon^{N+1}$, where $N$ is the normalization order.

After describing the computation of the proper elements in
Section~\ref{sec:NRR}, we consider three different samples at
increasing altitudes (see Sections~\ref{sec:ProperElementsLow},
\ref{sec:ProperElementsMed});
these samples are composed by a number of fragments generated after a break-up event,
which is obtained through the simulator briefly described in Section~\ref{sec:simulator} and
developed in \cite{simulator}, using the procedure presented in \cite{Johnson2001}.

The distribution of the proper elements at different
times of the evolution of the propagation of the fragments will be presented later
in Section~\ref{sec:ConstancyProperElements}.

\subsection{Computation of the proper elements} \label{sec:NRR}
We consider objects belonging to regions not affected by tesseral resonances,
so that the normalization procedure described in Section~\ref{sec:Normalization} applies straightforwardly.
For the elements of Sun and Moon, we will use the data given in Table~\ref{tab:SunMoon}.
The results of this Section concern break-up events and their propagation obtained by
considering the model including the geopotential, Sun and Moon. Hence, in this
Section the normalization procedure does not include SRP, which will be discussed in
Section~\ref{sec:SRP}, where we will mention the modifications needed in the normalization
to include the effect of SRP.

\vskip.1in

\begin{table}
    \begin{center}
\begin{tabular}{|c|c|c|}
  \hline
   & Sun & Moon \\
  \hline
  Mean daily motion & $1^o/day$ & $13.06^o/day$ \\
  Semi-major axis & $1.496\cdot 10^8$ km & $384478$ km \\
  Eccentricity & 0.0167 & 0.0549 \\
  Inclination & $23^o26'21.406''$ & $5^o15'$ \\
  $\dot\omega_{S/M}$ & $282.94^o/day$ & $0.164^o/day$ \\
  $\dot\Omega_{S/M}$ & $0^o/day$ & $-0.0529918^o/day$ \\
  \hline
\end{tabular}
\vskip.1in
        \caption{Orbital elements of Sun and Moon.}
\label{tab:SunMoon}
    \end{center}
\end{table}

\vskip.1in

The first step consists in averaging the expansions (\ref{eq:EarthExp}), (\ref{eq:MoonExp}), (\ref{eq:SunExp})
(truncated to order $l=2$)
over the fast and semi-fast angles, namely the mean anomalies $M$, $M_S$, $M_M$, and the sidereal time $\theta$.
As a consequence of the averaging over $M$, the semi-major axis is constant for the approximate
averaged Hamiltonian and becomes the first proper element.

Due to the fact that the quantity $\Omega_M$ depends on time with constant rate
$\dot\Omega_M=-0.0529918^o/day$ (as shown in Table~\ref{tab:SunMoon}), after averaging we end-up
with the following Hamiltonian function:
$$
{\overline{\mathcal{H}}}(e,i,\omega,\Omega,t) = \mathcal{H}_{E}^{sec}(e,i,\omega,\Omega) +
{\overline{\mathcal{H}}}_S(e,i,\omega,\Omega) + {\overline{\mathcal{H}}}_M(e,i,\omega,\Omega,t)\ ,
$$
where $\mathcal{H}_{E}^{sec}$ is given by \equ{eq:HamJ3Averaged} and includes terms depending on $J_2$ and $J_3$.

Let us introduce the Delaunay action variables defined as
$$
L=\sqrt{\mu_E a}\ ,\qquad G=L\sqrt{1-e^2}\ ,\qquad H=G\cos i\ ;
$$
the conjugated angle variables are $M$, $\omega$, $\Omega$. In terms of such variables,
we get a two degrees of freedom, time-dependent Hamiltonian of the form:

$$
{\overline{\mathcal{H}}}(G,H,\omega,\Omega,t) = \mathcal{H}_{E}^{sec}(G,H,\omega,\Omega) +
{\overline{\mathcal{H}}}_S(G,H,\omega,\Omega) + {\overline{\mathcal{H}}}_M(G,H,\omega,\Omega,t)\ .
$$

In the light of the theory presented in Section~\ref{sec:Normalization},
it is convenient to transform the Hamiltonian so that it becomes autonomous.
To do this, we premise that our unit of time is one sidereal day
over $2\pi$ and that the angles are measured in radians; since in
Table~\ref{tab:SunMoon} the rate of variation of the longitude of the ascending node
is given in degrees/synodic day, we transform such value as
$-0.0529918\cdot 365.242196/(366.242196\cdot 360)$, which gives the value of
$-0.000146798$ rad/(sidereal day). This motivates the introduction of
an angle variable defined as $\Omega_M=-0.000146798\, t$, which represents the linear evolution of the node
of the Moon in our time units; we denote by $H_M$ its conjugated action variable.
Then, the extended Hamiltonian is given by
\beqano
{\overline{\mathcal{H}}}_{ext}(G,H,H_M,\omega,\Omega,\Omega_M) &=& \mathcal{H}_{E}^{sec}(G,H,\omega,\Omega) +
{\overline{\mathcal{H}}}_S(G,H,\omega,\Omega)\nonumber\\
& + & {\overline{\mathcal{H}}}_M(G,H,\omega,\Omega,\Omega_M)-0.000146798\, H_M\ .
\eeqano
Next, we make a linear (canonical) change of coordinates to consider the dynamics
around fixed reference values of $G$ and $H$, say $G_0$ and $H_0$. Thus, the
transformation of coordinates $(G,H)\to(P+G_0,Q+H_0)$ leads
to introduce new variables $(P,Q)$, which are close to zero.
To keep a consistent notation, we use $(Q_M,p,q,q_M)$ instead of $(H_M,\omega,\Omega,\Omega_M)$. In
this notation, in the neighborhood of $G_0$, $H_0$, the Hamiltonian function can be written as

\beqano
{\overline{\mathcal{H}}}_{ext}(P+G_0,Q+H_0,Q_M,p,q,q_M) &=& \mathcal{H}_{E}^{sec}(P+G_0,Q+H_0,p,q) +
{\overline{\mathcal{H}}}_{S}(P+G_0,Q+H_0,p,q)\nonumber\\
&+& {\overline{\mathcal{H}}}_{M}(P+G_0,Q+H_0,p,q,q_M)-0.000146798\, Q_M\ ,
\eeqano

that we expand in power series around $P = 0$ and $Q = 0$ up to order 3 in $P$ and $Q$, separately.
We call ${\overline{\mathcal{H}}}$ the expanded Hamiltonian.

The linear part in the action variables of the Hamiltonian, that we denote by
$Z_0$, can be expressed in the form:
$$
Z_0(P,Q,Q_M)=\nu_P\, P+\nu_Q\, Q+\nu_{Q_M}\, Q_M\ ,
$$
where the `frequencies' $\nu_P$, $\nu_Q$, $\nu_{Q_M}$ are functions of $L_0$,
$G_0$, $H_0$. The explicit expressions for the frequencies are the following:
\beqano
\nu_P(L_0,G_0,H_0) &=& \frac{0.92\cdot 10^{-4} H_0^2}{G_0^6 L_0^3}-\frac{0.18\cdot 10^{-4}}{G_0^4
L_0^3}+\frac{0.33\cdot 10^{-4} H_0^2 L_0^4}{G_0^3}-0.67\cdot 10^{-4} G_0 L_0^2\ ,\nonumber\\
\nu_Q(L_0,G_0,H_0) &=& -\frac{0.37\cdot 10^{-4} H_0}{G_0^5 L_0^3}-\frac{0.33\cdot 10^{-4} H_0
L_0^4}{G_0^2}+0.20\cdot 10^{-4} H_0 L_0^2\ ,\nonumber\\
\nu_{Q_M} &=& -1.46798\cdot 10^{-4}\ .
\eeqano
We remind that the
units of length and time are normalized by setting the
geostationary distance of $42164.1696$ km equal to one and the
period of Earth's rotation equal to $2\pi$; this choice implies
$\mu_E = 1$.

Next, we split the Hamiltonian into two parts, namely $Z_0$ and a remainder $R_0$.
In this way we end up with a Hamiltonian of the form (\ref{eq:HamGenInit}), namely
\beq{HZR}
\mathcal{H}(P,Q,Q_M,p,q,q_M) = Z_0(P,Q,Q_M) + R_0(P,Q,Q_M,p,q,q_M)\ .
\eeq

Given that the Hamiltonian in \equ{HZR} is obtained as the sum of two contributions
with the norm of $R_0$ typically (much) smaller than the norm of $Z_0$,
following \cite{christos2012book}, we introduce a book-keeping parameter $\lambda$
in front of $R_0$.
More precisely, the book-keeping parameter is introduced to label
the part of the Hamiltonian which has to be removed by the normalization algorithm.
Therefore, following the procedure described in Section~\ref{sec:Normalization}, at each
normalization step we split the remainder into two parts, the first one depending just on the actions and
the second one depending on all variables. Performing the steps
(\ref{eq:HomologicGen})-(\ref{eq:CoefChi}), we compute the generating function and the transformed
Hamiltonian.

If we stop the iteration of the normalization procedure, we retain the Hamiltonian parts
containing only the terms independent on $\lambda$ and the terms linear in $\lambda$.
Since the introduction of the book-keeping was fictitious, at the end we restore
its value to $\lambda=1$.

If, instead, we decide to perform another normalization step, the terms of
second order in $\lambda$ are used to label the remainder and the normalization method
is then iterated. Once we finish the iteration, we use the generating
functions computed at each normalization step to find the other two proper elements, namely
the proper eccentricity and the proper inclination.

To be more specific, let us denote by $N$ the number of steps of the normalization procedure
(in practical computations we will take $N=1$). Denoting with a prime the variables after the
normalizing transformation, we end-up with a Hamiltonian of the form
\beq{HN}
{\mathcal{H}}^{(N)}(P',Q',Q_M',p',q',q_M')=Z^{(N)}(P',Q',Q_M')+R^{(N)}(P',Q',Q_M',p',q',q_M')\ ,
\eeq
where $Z^{(N)} = Z_0(P',Q',Q_M') + \lambda Z_1(P',Q',Q_M') + \dots + \lambda^N Z_N(P',Q',Q_M')$ is the normal form at order $N$, depending just on the actions,
and $R^{(N)} = \lambda^{N+1} R_N(P',Q',Q_M',p',q',q_M')$ is the remainder; again, $\lambda$ denotes the book-keeping parameter,
that will be set equal to one at the end of the procedure. If we disregard $R^{(N)}$, we obtain that
$P'$, $Q'$, $Q_M'$ are constants of motion for the Hamiltonian
${\mathcal{H}}_0^{(N)}=Z^{(N)}(P',Q',Q_M')$, while $p'$, $q'$, $q_M'$ evolve linearly in time. Hence, we can write
the solution of Hamilton's equations associated to ${\mathcal{H}}_0^{(N)}$ as
\begin{align*}
P'(t) &= P_0' \\
Q'(t) &= Q_0'\\
Q_M'(t) &= Q_{M,0}'\\
p'(t) &= p_0' + \dfrac{\partial Z^{(N)}}{\partial P'}(P'_0,Q'_0,Q_{M,0}')\ t \\
q'(t) &= q_0'+ \dfrac{\partial Z^{(N)}}{\partial Q'}(P'_0,Q'_0,Q_{M,0}')\ t \\
q_M'(t) &= q_{M,0}'+ \dfrac{\partial Z^{(N)}}{\partial Q_M'}(P'_0,Q'_0,Q_{M,0}')\ t\ ,
\end{align*}
where $P'_0,Q'_0,Q_{M,0}',p_0',q_0',q_{M,0}'$ denote the initial conditions.
To compute the original variables as a function of the new variables, we use the generating functions that
led to \equ{HN} and that we denote as $\chi^{(N)},\chi^{(N-1)},\dots,\chi^{(1)}$ (\cite{Deprit69}). Hence, we obtain
\begin{align*}
P(P'(t),Q'(t),Q_M'(t),p'(t),q'(t),q_M'(t))=S^\lambda_{\chi^{(1)}}\circ\dots\circ S^\lambda_{\chi^{(N)}}P',\\
Q(P'(t),Q'(t),Q_M'(t),p'(t),q'(t),q_M'(t))=S^\lambda_{\chi^{(1)}}\circ\dots\circ S^\lambda_{\chi^{(N)}}Q',\\
Q_M(P'(t),Q'(t),Q_M'(t),p'(t),q'(t),q_M'(t))=S^\lambda_{\chi^{(1)}}\circ\dots\circ S^\lambda_{\chi^{(N)}}Q_M',\\
p(P'(t),Q'(t),Q_M'(t),p'(t),q'(t),q_M'(t))=S^\lambda_{\chi^{(1)}}\circ\dots\circ S^\lambda_{\chi^{(N)}}p',\\
q(P'(t),Q'(t),Q_M'(t),p'(t),q'(t),q_M'(t))=S^\lambda_{\chi^{(1)}}\circ\dots\circ S^\lambda_{\chi^{(N)}}q',\\
q_M(P'(t),Q'(t),Q_M'(t),p'(t),q'(t),q_M'(t))=S^\lambda_{\chi^{(1)}}\circ\dots\circ S^\lambda_{\chi^{(N)}}q_M'\ ,
\end{align*}
which implies that we can express the original variables as
\begin{align}\label{eq:solutionOldVarParam}
P(P'(t),Q'(t),Q_M'(t),p'(t),q'(t),q_M'(t))=\tilde{P}(P_0',Q_0',Q_{M,0}',p_0',q_0',q_{M,0}',t),\\ \nonumber
Q(P'(t),Q'(t),Q_M'(t),p'(t),q'(t),q_M'(t))=\tilde{Q}(P_0',Q_0',Q_{M,0}',p_0',q_0',q_{M,0}',t),\\ \nonumber
Q_M(P'(t),Q'(t),Q_M'(t),p'(t),q'(t),q_M'(t))=\tilde{Q}_M(P_0',Q_0',Q_{M,0}',p_0',q_0',q_{M,0}',t),\\ \nonumber
p(P'(t),Q'(t),Q_M'(t),p'(t),q'(t),q_M'(t))=\tilde{p}(P_0',Q_0',Q_{M,0}',p_0',q_0',q_{M,0}',t),\\ \nonumber
q(P'(t),Q'(t),Q_M'(t),p'(t),q'(t),q_M'(t))=\tilde{q}(P_0',Q_0',Q_{M,0}',p_0',q_0',q_{M,0}',t),\\ \nonumber
q_M(P'(t),Q'(t),Q_M'(t),p'(t),q'(t),q_M'(t))=\tilde{q}_M(P_0',Q_0',Q_{M,0}',p_0',q_0',q_{M,0}',t)\ , \nonumber
\end{align}
where, by using the solutions of the normal form equations, the functions
$\tilde{P}$, $\tilde{Q}$, $\tilde{Q}_M$, $\tilde{p}$, $\tilde{q}$, $\tilde{q}_M$ are
explicit functions of time.
To compute $P_0',Q_0',Q_{M,0}',p_0',q_0',q_{M,0}'$, we have two possibilities. One is to compute
the inverse transformation by exploiting the generating functions;
however, for a low order normalization, this procedure is usually not very accurate.
On the other hand, an appropriate method to compute the initial conditions is to solve the following system of equations:
\begin{align*}
\tilde{P}(P_0',Q_0',Q_{M,0}',p_0',q_0',q_{M,0}',0)=P_0,\\
\tilde{Q}(P_0',Q_0',Q_{M,0}',p_0',q_0',q_{M,0}',0)=Q_0,\\
\tilde{Q}_M(P_0',Q_0',Q_{M,0}',p_0',q_0',q_{M,0}',0)=Q_{M,0},\\
\tilde{p}(P_0',Q_0',Q_{M,0}',p_0',q_0',q_{M,0}',0)=p_0,\\
\tilde{q}(P_0',Q_0',Q_{M,0}',p_0',q_0',q_{M,0}',0)=q_0,\\
\tilde{q}_M(P_0',Q_0',Q_{M,0}',p_0',q_0',q_{M,0}',0)=q_{M,0}\ ,
\end{align*}
which we insert in \equ{eq:solutionOldVarParam} to obtain the
analytic solution in the original variables.
Finally, from $P(t)$, $Q(t)$, we go back to $G(t)$, $H(t)$ and hence to $e(t)$, $i(t)$.
As a last step, we define the proper eccentricity $e_p$ and the proper inclination $i_p$
as the averages over a given time interval, say $[t_0,T]$:
\begin{align*}
e_p(T)=\frac{1}{T-t_0}\int_{t_0}^{T} e(t) dt, \qquad
i_p(T)=\frac{1}{T-t_0}\int_{t_0}^{T} i(t) dt \ .
\end{align*}
We remark that we could have also computed the proper elements in the transformed variables,
without the need of using the analytic solution to compute the proper elements in
terms of the original variables. Having in mind concrete applications, we believe
that our procedure is more appropriate, beside being suitable for different purposes,
like the propagation of the solution. However, it must be admitted that the direct determination
of the proper elements in the transformed variables is computationally simpler and it provides results quite
similar to our original procedure.

\subsection{A simulator for generating fragments}\label{sec:simulator}
In the following Sections, we will compute the proper elements associated to several fragments generated
by a break-up event. To this end, our results are obtained using a simulator of collisions developed
within the ongoing collaboration in \cite{simulator}, which reproduces the break-up model Evolve 4.0 provided by NASA (see
\cite{Johnson2001}, \cite{Klinkrad}, \cite{Report}). This simulator, which has
been validated for debris sizes greater than 1 mm, allows us to determine the cross-sections,
masses, and imparted velocities of the fragments after an explosion or a collision.

The break-up model Evolve 4.0 includes:

- the size distribution of the fragments after collision or explosion,

- the fragments' area-to-mass ratio,

- the fragments' relative velocity distribution with respect to the parent body.

Due to the fact that the above parameters are not the same for all debris,
it is necessary to provide the distributions as a function of a given parameter,
e.g. the mass or the characteristic length. Besides, the simulations can be highly influenced by the
initial conditions and parameters of the break-up, for example
the total mass of the parent body or the collision velocity.

We remark that the explosion and collision rates, as well as the fragment size distribution,
affect the production rate of the debris particles.
On the other hand, the area-to-mass ratio and the relative velocity distribution
affect how the debris evolve and eventually decay.

We refer to \cite{Johnson2001}, \cite{Klinkrad},
\cite{Report} for further details on the break-up
simulator that we reproduced in \cite{simulator} to obtain the fragments generated by
a collision or an explosion.

In the collision case, the simulator
has as inputs the position of the parent body, the mass of both
parent and projectile bodies and the impact velocity.
In the explosion case, the inputs are just the position and type of the parent satellite (body).

Only fragments bigger than 12 cm are generated. The output that will be produced to  obtain
the results of Sections~\ref{sec:ProperElementsLow}-\ref{sec:grouping} is a combination of the
simulator of break-up events, the propagation of the fragments' orbits, and the computation of
the proper elements. Precisely, we compute the following three sets of data:

$(i)$ a set of instantaneous
velocities after break-up, transformed to the orbital elements of
each fragment at the instant of time after the event. In this way, we
obtain the distribution of the elements in the phase space of
semi-major axis, eccentricity and inclination;

$(ii)$ starting from the data after the break-up, we
propagate each fragment for a given period of time, typically up to
150 years. This result allows us to monitor how the distribution evolves;

$(iii)$ we use the position of the fragments after the propagation up to a given interval of time
(e.g., 150 years) to
compute the proper elements of each fragment. The distribution of the
proper elements is then compared to that of the elements at the initial time
after the break-up.

\vskip.1in

Among an extensive survey of cases we have analyzed, we select the three representative
samples at different altitudes described in Sections~\ref{sec:ProperElementsLow}, \ref{sec:ProperElementsMed}.

\subsection{Moderate altitude orbits}\label{sec:ProperElementsLow}
In this Section we present two samples obtained by simulating an explosion of a spacecraft of {\sl Titan Transtage} type and a collision between a 1200 kg parent body and 5 kg projectile at the relative velocity of 4900 m/s.

\begin{figure}[h]
\centering
\includegraphics[width=0.8\columnwidth]{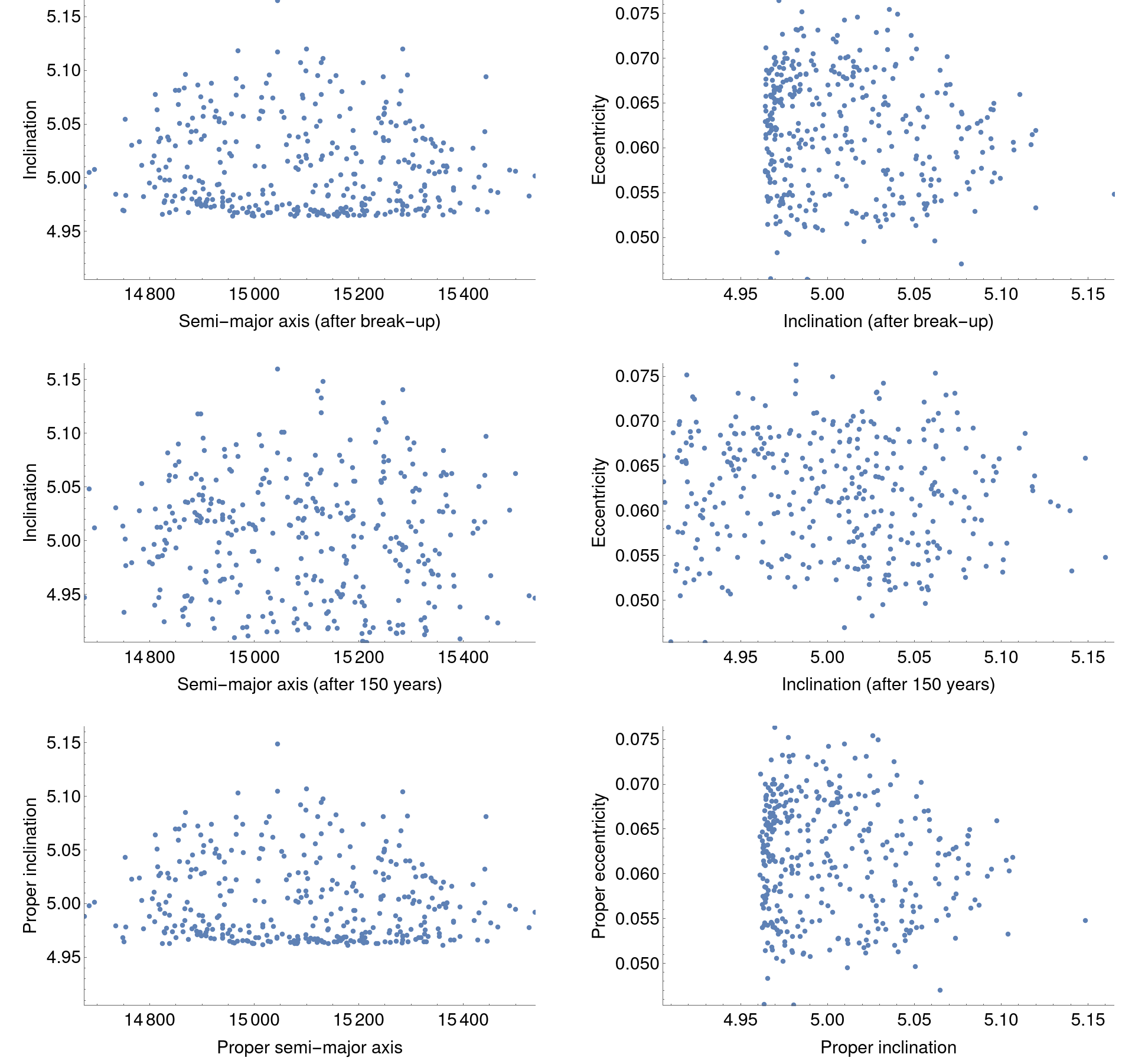}
\caption{Distribution of $a$-$i$ (left), $i$-$e$ (right) with the parent body at the
initial position $a=15100$ km, $e=0.06$, $i=5^o$, $\omega = 90^o$, $\Omega = 10^o$ after break-up (first row), after 150 years
(second row), and proper elements computed after 150 years (third
row).} \label{fig:ModAt1FullRange}
\end{figure}

The first sample is located at relatively low distance from the Earth; precisely,
we consider a simulated explosion that generates 356 fragments and having
the following parameters of the parent body: $a=15100$ km, $e=0.06$, $i=5^o$, $\omega = 90^o$, $\Omega = 10^o$.
The different panels of Figure~\ref{fig:ModAt1FullRange}
show the osculating elements immediately after the break-up (first row),
the mean elements (obtained by integrating the averaged Hamiltonian) propagated up to 150 years (second row)
and the proper elements
computed from the evolution after 150 years (third row). The reason for
computing the proper elements after a time interval (say, 150 years) relies on
the fact that space debris catalogues, like the TLE (Two-Line-Elements) catalogue, provide
the values of the elements at a given epoch after the disruptive event and usually not at the time at which
the break-up takes place.

Figure~\ref{fig:ModAt1FullRange} illustrates the results in the planes $a$-$i$, $i$-$e$, which show that
the proper elements give important information on the distribution
of the fragments after the collision. Indeed,
after 150 years the debris are spread in phase space (second row), while
in the proper element phase space the fragments are restored in groups (third row),
similar to those obtained just after break-up (first row).
We notice a small shift of the proper inclination, as it happens in most of the examples
described in the rest of the work; sometimes the shift occurs also in the eccentricity.

\begin{figure}[h]
\centering
\includegraphics[width=0.8\columnwidth]{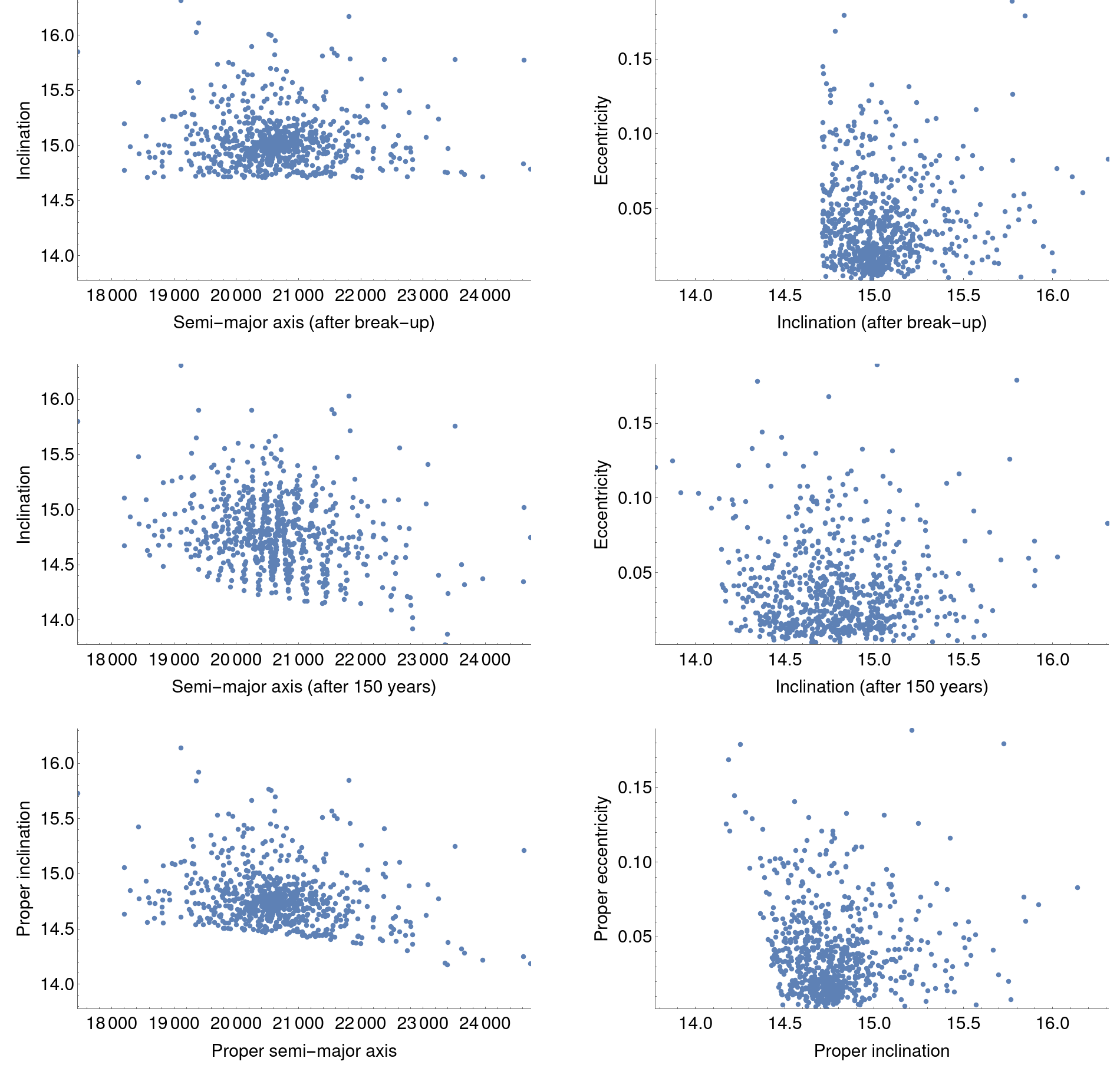}
\caption{Distribution of $a$-$i$ (left), $i$-$e$ (right) with the parent body at the initial position $a=20600$ km,
$e=0.01$, $i=15^o$, $\omega = 10^o$, $\Omega = 20^o$ after break-up (first row),
after 150 years (second row), and proper elements computed after 150 years (third row).}
\label{fig:ModAt2FullRange}
\end{figure}

The second sample concerns a collision that generates 767 fragments. The event occurs at a moderate altitude of the parent body with
$a=20600$ km and with a relatively small inclination and eccentricity,
$e=0.01$, $i=15^o$; the other elements are fixed as
$\omega = 10^o$ and $\Omega = 20^o$.

Figure~\ref{fig:ModAt2FullRange} gives the results in the planes $a$-$i$ and $i$-$e$,
where the scales have been fixed as the minimum and the maximum values of the
evolution of the elements after 150 years.

A comparison between the panels of the first and second row of Figure~\ref{fig:ModAt2FullRange}
shows that the fragments are moderately sparse after a propagation up to 150 years; on the other hand, the
proper elements plots look similar to the distribution after the break-up,
thus allowing to establish a connection with the fragments at the break-up event.

\begin{figure}[h]
\centering
\includegraphics[width=1\columnwidth]{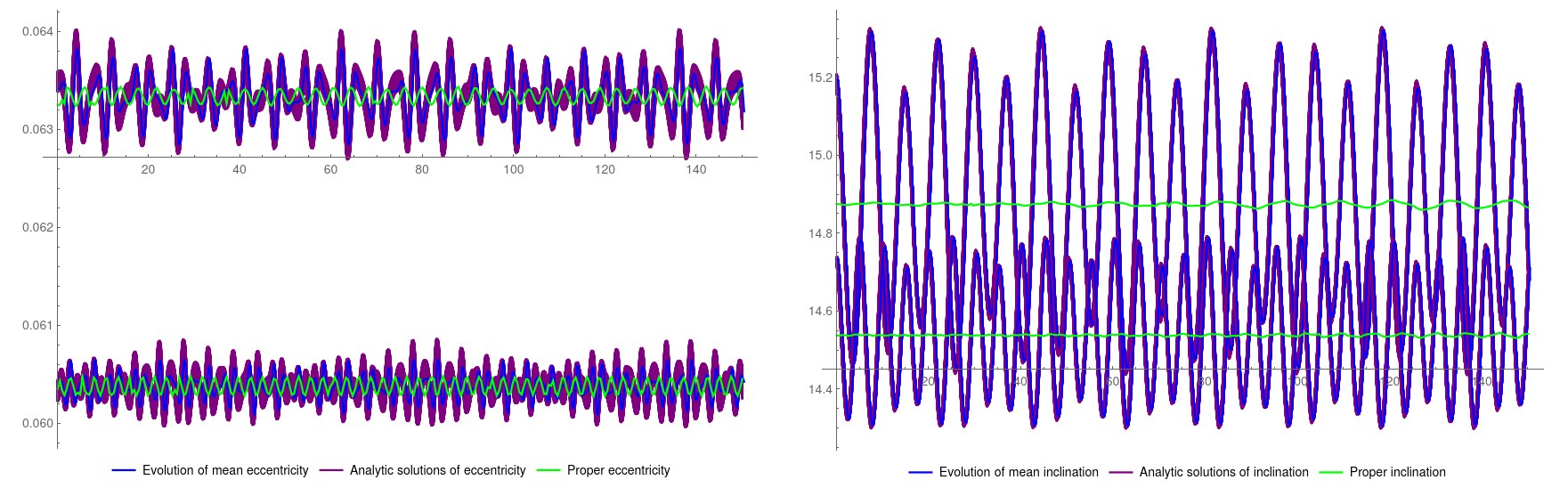}
\caption{Proper values and evolution of eccentricity and inclination
over 150 years of two random fragments ($a=20600$ km, $e=0.01$, $i=15^o$,
$\omega = 10^o$, $\Omega = 20^o$). The blue line is the evolution
of the mean elements; the purple line is the analytic solution computed every 6 months;
the green line denotes the proper element computed every 6 months.}
\label{fig:ModAt2SingEV}
\end{figure}

If we get a closer look at two random fragments as shown in Figure~\ref{fig:ModAt2SingEV},
we find that the proper values for the orbital elements are very close to
the average values of the evolution of the osculating elements. The
evolution of the inclination shows that at a given time the two
fragments can be apart by about $1^o$, while the proper elements
(given by the average inclinations of both fragments) are closer, but
nearly constant.

\subsection{Medium altitude orbits}  \label{sec:ProperElementsMed}
Next case concerns a sample located in the GEO region, between the resonances 2:1
and 1:1. The event has been simulated taking a parent body at
$a= 33600$ km, $e=0.05$, $i=20^o$, $\omega = 120^o$, $\Omega = 60^o$
and it consists of an explosion of a spacecraft of {\sl Titan Transtage} type.
After break-up, a total of 356 fragments have been generated.

\begin{figure}[h]
\centering
\includegraphics[width=0.8\columnwidth]{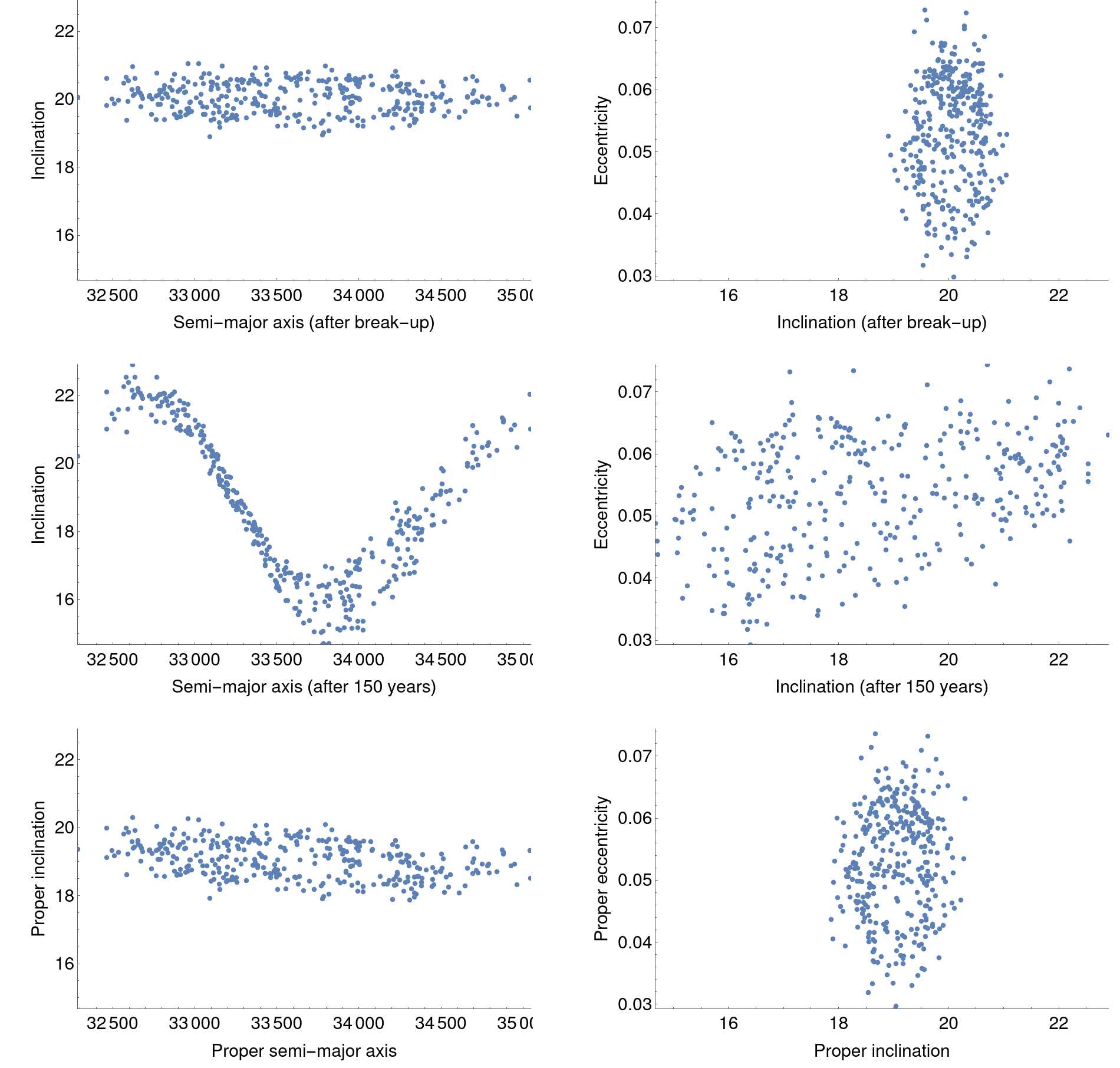}
\caption{Distribution of $a$-$i$ (left), $i$-$e$ (right) with the parent body at the initial position
$a= 33600$ km, $e=0.05$, $i=20^o$, $\omega = 120^o$, $\Omega = 60^o$ after break-up (first row),
after 150 years (second row), and proper elements computed after 150 years (third row).}
\label{fig:MidAt2FullRange}
\end{figure}

The results providing the elements at break-up, the propagation after 150 years and
the proper elements computed using the data after 150 years are given
in Figure~\ref{fig:MidAt2FullRange}.
The evolution of the osculating elements after 150 years in the $a$-$i$ plane
seems to be located around a fitting parabolic curve; instead,
the fragments are definitely sparse in the $i$-$e$ plane after 150 years, due
to a growth of the inclination from about $15^o$ to $22^o$.
On the contrary, the proper elements in the $a$-$i$ plane take a shape similar
to the distribution at the break-up around the value $i=20^o$.

\subsection{Identification through fragments' grouping}\label{sec:grouping}
\begin{figure}[h]
\centering
\includegraphics[width=0.8\columnwidth]{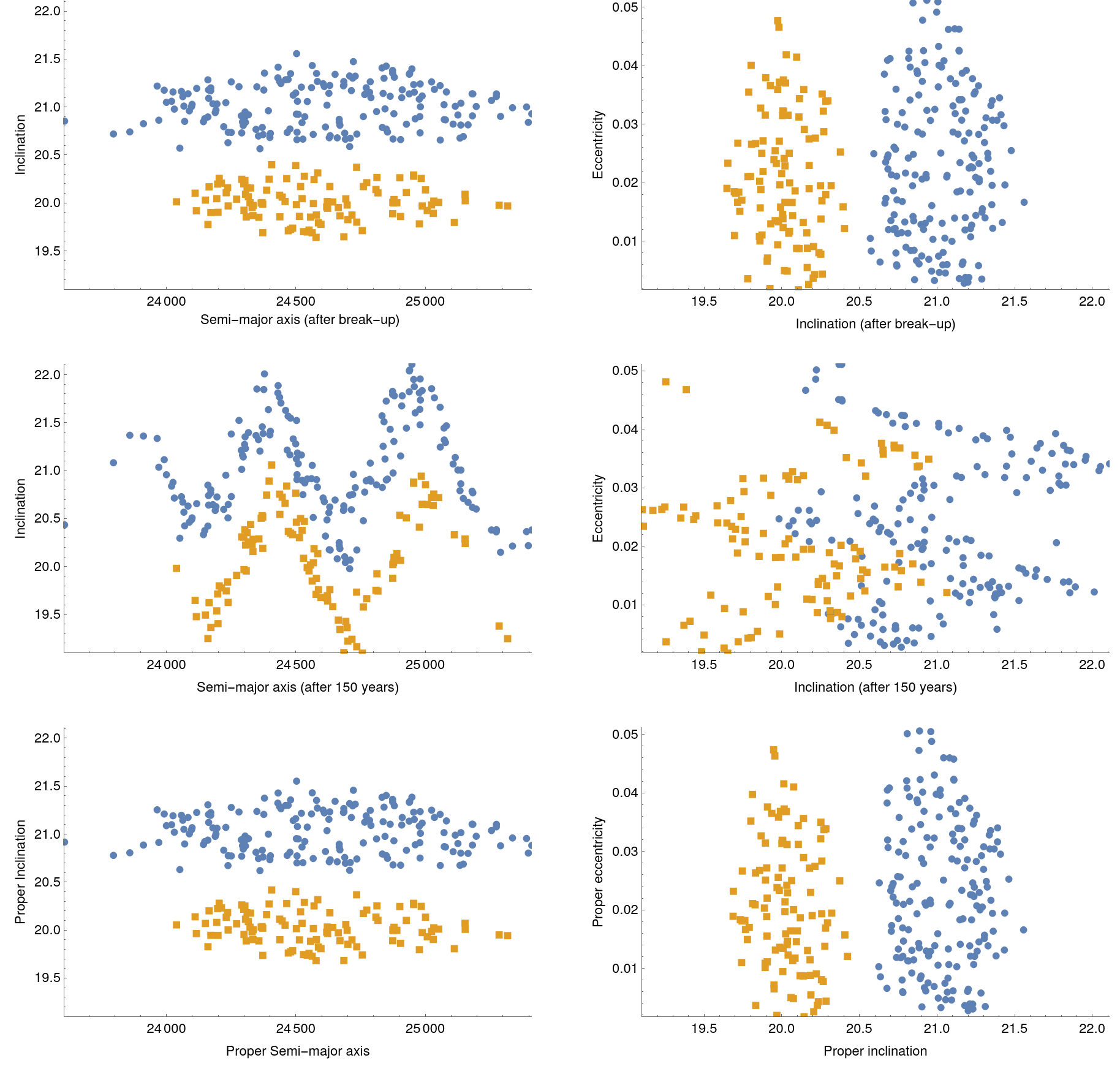}
\caption{Distribution of $a$-$i$ (left), $i$-$e$ (right) with the parent body at the initial position $a= 24600$ km,
$e=0.02$, $i=20^o$ and $i=21^o$, $\omega = 110^o$, $\Omega = 120^o$ at break-up (first row), after 150 years (second row) and proper elements
computed from data after 150 years (third row).}
\label{fig:groups2}
\end{figure}

An important {\sl practical} use of the proper elements computation might come
from the classification of events taking place at nearby
locations. To this end, we simulate in Figure~\ref{fig:groups2}
two separate explosions having all elements in common, except the
inclination: one explosion occurs at $i=20^o$ and the other at $i=21^o$.
While the evolution after 150 years does not permit to distinguish between
the two groups belonging to the original parent bodies, the computation
of the proper elements allows us to distinguish two clear groups that
closely resemble the distribution of fragments after the break-up.

\begin{figure}[h]
\centering
\includegraphics[width=0.8\columnwidth]{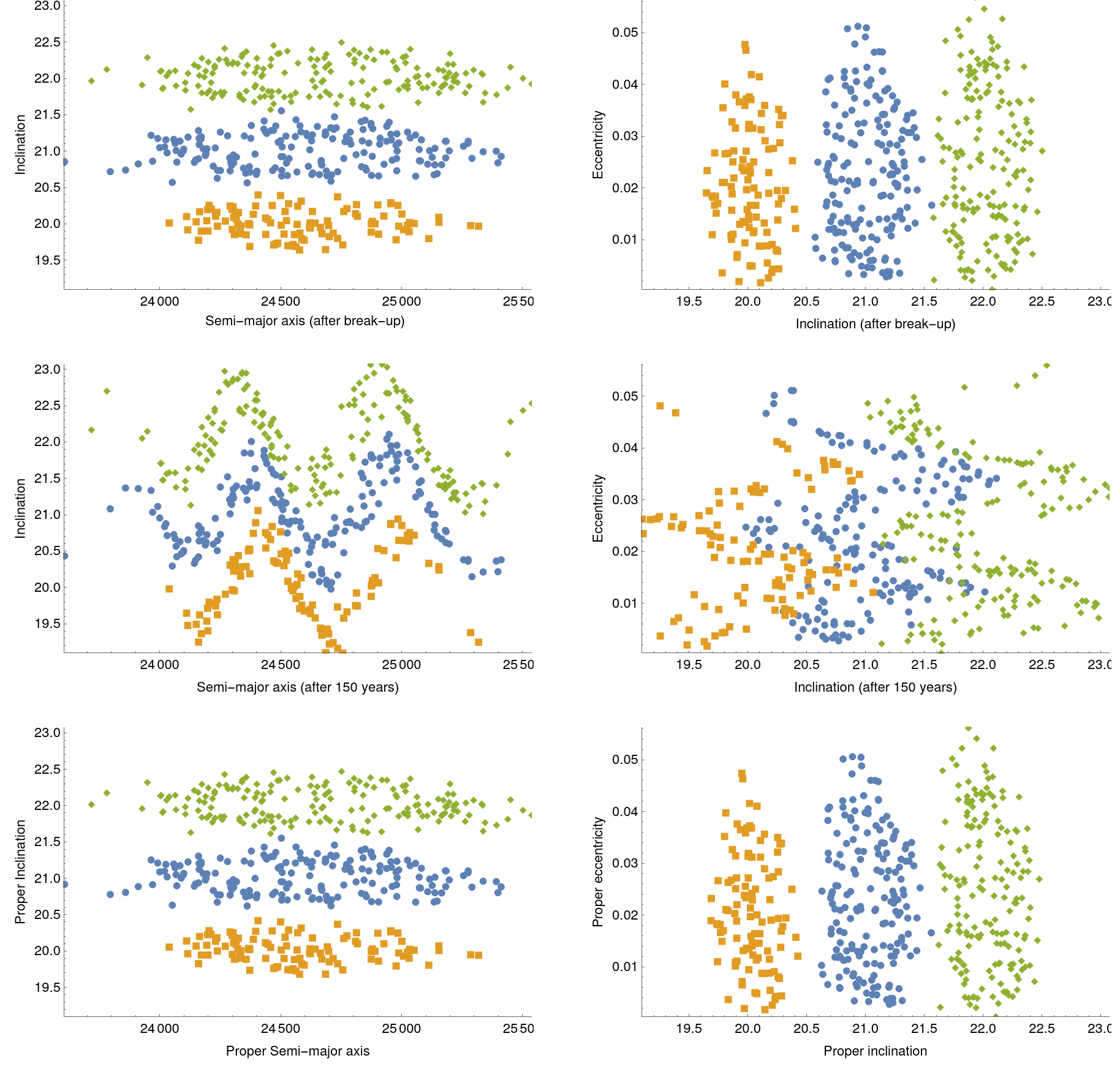}
\caption{Distribution of $a$-$i$ (left), $i$-$e$ (right) with the parent body at the initial position $a= 24600$ km,
$e=0.02$, $i=20^o$, $i=21^o$, $i=22^o$, $\omega = 110^o$, $\Omega = 120^o$ at break-up (first row), after 150 years (second row) and proper elements
computed from data after 150 years (third row).}
\label{fig:groups3}
\end{figure}

We also provide a simulation of three explosions occurring at three different inclinations, precisely
$i=20^o$, $i=21^o$, $i=22^o$. Also in this case, see Figure~\ref{fig:groups3}, the proper elements
computation allows us to get information on the existence of three different groups, which correspond
to the three different explosions.

These two simple examples confirm the validity of the computation of the proper elements for space debris, as already witnessed
by the results on the identification of asteroid families;
the formation of groups, as well as their resemblance with the dynamical plots after the break-up,
might be used to reconstruct the catastrophic event and to identify the origin of the fragments.

\section{Proper elements close to a tesseral resonance}\label{sec:ProperElementsTes21}
In this Section, we focus on break-up events taking place close to tesseral
resonances. The procedure to construct the normal form will be
the same as in Section~\ref{sec:ProperElements}, since our aim is to see if the computation of the proper
eccentricity and the proper inclination can be still used to
reconstruct the initial distribution after the break-up event.

Tesseral resonances have a strong influence on the evolution of semi-major axis, while
their effect is less important for inclination and eccentricity.
We include the effect of the tesseral resonance in the computation of
the evolution as described in the following Sections.

\subsection{Close to the 2:1 tesseral resonance} \label{sec:ProperElementsTes11}
We start by analyzing the behavior close to the 2:1 tesseral resonance.
We remind that the 2:1 tesseral resonance occurs whenever there is a commensurability relation
between the mean motion of the space debris, the sidereal time, the argument of perigee and
the longitude of the ascending node:
$$
\dot{M}-2\dot{\theta}+2\dot{\Omega}+\dot{\omega}=0\ .
$$
We remark that when $J_2=0$, then $\dot\omega=\dot\Omega=0$ and the resonance relation reduces to
$$
\dot M-2\dot\theta=0\ .
$$
A {\sl multiplet} tesseral resonance (see \cite{cellettietal2020}) occurs whenever the following relation is satisfied for
$\ell\in\integer\backslash\{0\}$:
$$
\dot{M}-2\dot{\theta}+2\dot{\Omega}+\dot{\omega}+\ell\dot\omega=0\ .
$$
For the 2:1 resonance, we consider a Hamiltonian function composed by the following terms: the Keplerian
part, the resonant Hamiltonian limited to the most important terms, the secular Hamiltonian, the contributions
of Sun and Moon that we limit to the Hamiltonians averaged over the corresponding satellite mean anomaly. With respect
to the latter choice, we remark that we could have inserted the complete Sun and Moon Hamiltonians, but we noticed
that the non-average terms do not contribute much to the dynamics, while they remarkably increase
the complexity of the normalization procedure. In summary, in the proximity of the 2:1 resonance, we consider the
following Hamiltonian:
\beqano
\mathcal{H}(L,G,H,M,\omega,\Omega,\theta,t) &=& \mathcal{H}_{Kep}(L)+\mathcal{H}_{E}^{res2:1}(L,G,H,M,\omega,\Omega,\theta)
+ \mathcal{H}_{E}^{sec}(G,H,\omega,\Omega)\nonumber\\
&+& {\overline{\mathcal{H}}}_{S}(G,H,\omega,\Omega) + {\overline{\mathcal{H}}}_{M}(G,H,\omega,\Omega,t)\ ,
\eeqano
where the expansion (in the Keplerian orbital elements)
of the term $\mathcal{H}_{E}^{res2:1}$ up to order $n=m=3$ and second order in eccentricity
(including multiplet resonant terms) is given by
\beqa{Hres21}
\mathcal{H}_{E}^{res2:1} &=& \frac{J_{22}  \mu _{E} R_{E}^2}{a^3} \left(\frac{9}{8} e \left(2-2 \cos ^2(i)\right)
\cos \left(-2 \lambda _{22}+2 \Omega - 2\theta +M\right)\right. \nonumber\\
&-&\left.\frac{3}{8} e \left(\cos ^2(i)+2 \cos (i)+1\right) \cos \left(-2 \lambda _{22}+2 \Omega - 2\theta
+M+2 \omega \right)\right)\nonumber\\
&+&\frac{J_{32} \mu _{E} R_{E}^3}{a^4} \left(\frac{165}{64} e^2  \sin (i)
\left(3 \cos ^2(i)-2 \cos (i)-1\right) \sin \left(-2 \lambda _{32}+2 \Omega - 2\theta +M-\omega \right)\right.\nonumber\\
&+& \left.\frac{15}{64} e^2 \sin (i) \left(\cos ^2(i)+2 \cos (i)+1\right)
\sin \left(-2 \lambda _{32}+2 \Omega - 2\theta +M+3 \omega \right)\right.\nonumber\\
&+&\left.\frac{15}{8} \left(2 e^2+1\right)\sin (i) \left(-3 \cos ^2(i)-2 \cos (i)+1\right)
\sin \left(-2 \lambda _{32}+2 \Omega - 2\theta +M+\omega \right)\right)\ .\nonumber\\
\eeqa
When the inclination is not too small, the largest term
in \equ{Hres21} is the last one containing the resonant angle,
which appears with the $J_{32}$ coefficient. On the other hand, as
shown in \cite{CellettiGales2014}, there are other two terms,
depending respectively on the combinations of the angles
$M-2\theta+2\Omega$ and $M-2\theta+2\Omega+2\omega$, which might
be important for specific values of eccentricity and inclination.
This leads to retain, in the first approximation, only three
terms in the initial Hamiltonian $\mathcal{H}_{E}^{res2:1}$. Hence,
the Hamiltonian used for the computation of the evolution of the
elements is the
following:

\beqano
&&\mathcal{H}(L,G,H,H_M,M,g,h,h_M,\theta) = -\frac{\mu_E^2}{2 L^2} + \frac{J_{2}
\mu_E^4 R_{E}^2 \left(G^2-3 H^2\right)}{4 G^5 L^3}
+ \mathcal{H}_{S}(G,H,g,h) \nonumber\\
&&+ \mathcal{H}_{M}(G,H,g,h,h_M) - 0.000146798H_M  + \frac{3}{64 G^2 L^{10}} \left(5 J_{32}
\mu_{E}^5 R_E^3  \right. \nonumber\\
&&\left. \sqrt{1-\frac{H^2}{G^2}} \left(-8 \left(G^2-2 G H-3 H^2\right) \left(2 G^2-3 L^2\right)
\sin \left(-2 \lambda_{32}+2 \Omega -2 \theta +M+\omega \right)\right)  \right. \nonumber\\
&&\left. +8 L^4 J_{22} \mu_{E}^4 R_{E}^2 \sqrt{1-\frac{G^2}{L^2}}
\left(6 (G^2-H^2) \cos(-2 \lambda_{22}+2 \Omega-2 \theta +M)  \right.\right. \nonumber\\
&&\left.\left. -(G+H)^2 \cos \left(-2\lambda_{22}+2\Omega-2\theta+M +2\omega  \right)\right)\right)\ .
\eeqano

\begin{figure}[h]
\centering
\includegraphics[width=0.8\columnwidth]{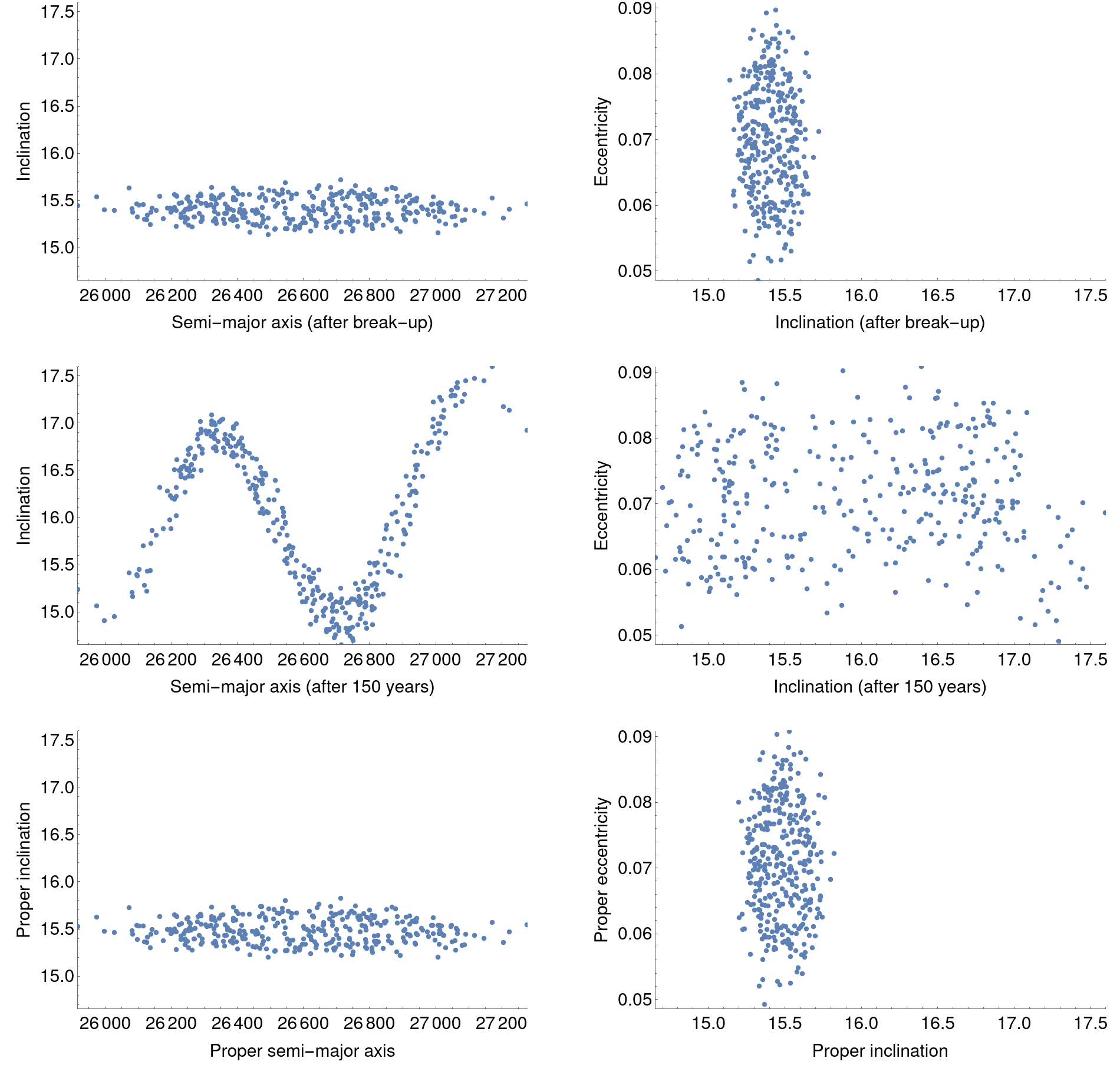}
\caption{Distribution of $a$-$i$ (left), $i$-$e$ (right) with the parent body at the initial position $a= 26600$ km, $e=0.07$, $i=15^o$,
$\omega = 270^o$, $\Omega = 120^o$ after break-up (first row), after 150 years (second row),
and proper elements computed after 150 years (third row).}
\label{fig:Res21At2FullRange}
\end{figure}

Figure~\ref{fig:Res21At2FullRange} shows an experiment analyzing
an explosion of a spacecraft of {\sl Titan Transtage} type
orbiting at $a= 26600$ km, $e=0.07$, $i=15^o$, $\omega = 270^o$,
$\Omega = 120^o$. The dynamics around the 2:1 tesseral resonance
is chaotic, as shown by the sinusoidal distribution in the $a$-$i$
plane and from the spreading of fragments in inclination in
Figure~\ref{fig:Res21At2FullRange}. On the other hand, looking at
the third row of Figure~\ref{fig:Res21At2FullRange}, we notice a quite successful
computation of proper elements for almost all fragments. However,
we are aware of the fact that we did not obtain an accurate reconstruction of the
initial distribution, due to the fact that some fragments could be
very close to the resonant region, either inside it. We believe
that the computation of the proper elements close to a resonance might be improved by
using an appropriate resonant normal form procedure.

\subsection{Close to the 1:1 tesseral resonance} \label{sec:ProperElementsTes11}
In this Section, we shortly analyze the dynamics close to the 1:1 tesseral
resonance, which contains many resonant terms, but only a single dominant one,
as described in \cite{CellettiGales2014}. The dominant term is
given by the expression:
$$
\mathcal{H}_E^{res1:1}  = {{J_{22} \mu_E R_E^2}\over a^3}
\left.\frac{3}{4} \left(1-\frac{5
e^2}{2}\right) \left(\cos ^2(i)+2 \cos (i)+1\right) \cos
\left(-2 \lambda _{22}+2 \Omega - 2\theta+2 M+2 \omega
\right)\right)\ ,
$$
which leads to consider the following Hamiltonian:
\beqano
&&\mathcal{H}(L,G,H,l,g,h,\theta,h_M,H_M) = -\frac{\mu_E^2}{2 L^2} +
\frac{J_{2} \mu _{E}^4 R_{E}^2 \left(G^2-3
H^2\right)}{4 G^5 L^3}  + \mathcal{H}_E^{res1:1} (L,G,H,l,g,h,\theta)\nonumber\\
&&\qquad+ \mathcal{H}_{Sun}(G,H,g,h) + \mathcal{H}_{Moon}(G,H,g,h,g_M) -
0.000146798\, H_M\ .
\eeqano

\begin{figure}[h]
\centering
\includegraphics[width=0.8\columnwidth]{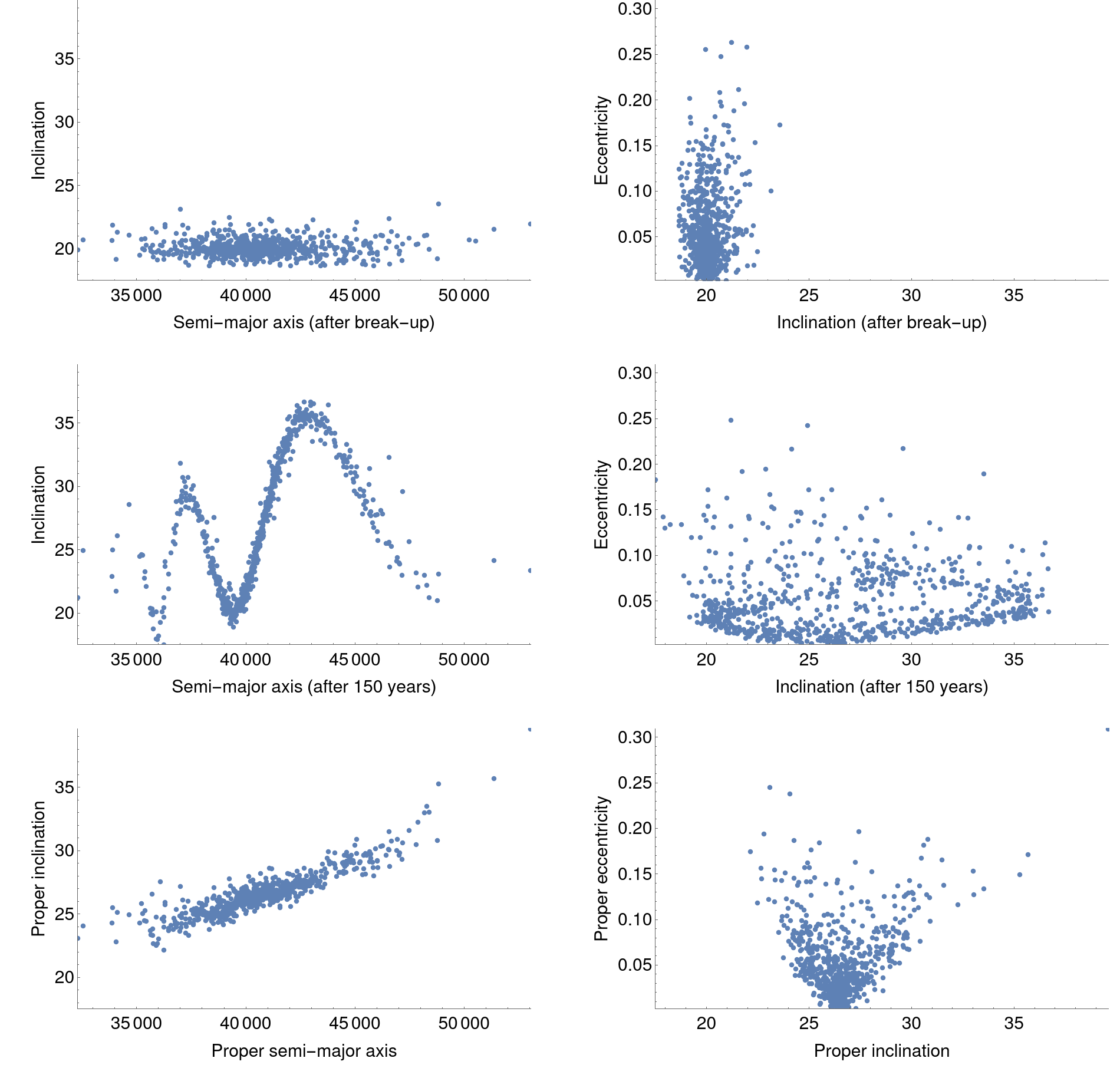}
\caption{Distribution of $a$-$i$ (left), $i$-$e$ (right) with the parent body at the initial position $a= 40600$ km, $e=0.01$, $i=20^o$, $\omega = 170^o$, $\Omega = 210^o$ after break-up (first row), after 150 years (second row),
and proper elements computed after 150 years (third row).}
\label{fig:Res11At2FullRange}
\end{figure}

Using the same approach as for the 2:1 resonance, we make
an experiment very close to the 1:1 resonance, taking $a= 40600$ km,
$e=0.01$, $i=20^o$, $\omega = 170^o$, $\Omega = 210^o$. In total,
815 fragments have been produced from a collision between a
1300 kg parent body and a 6 kg projectile at the relative velocity
of 4900 m/s.

The distributions
of proper eccentricity and proper inclination are closer to the
original distributions of the fragments, than the distributions
obtained using the mean elements. As we already mentioned for the 2:1 resonance, we believe that
the whole procedure requires more work and might be improved by using an
appropriate resonant normal form procedure.

\section{Data analysis, SRP, Noisy data, Constancy over time}
In this Section, we aim at supporting the results obtained in the
previous Sections by analyzing different aspects:

\begin{itemize}
    \item[$(i)$] we use statistical data analysis to quantify the
    similarities of the distributions of the fragments (see
    Section~\ref{sec:data});
    \item[$(ii)$] we analyze the effect of Solar radiation
pressure (see Section~\ref{sec:SRP});
    \item[$(iii)$] we make some experiments to study how the distributions are
    affected by noise (see Section~\ref{sec:RandomNoises});
    \item[$(iv)$] we provide some experiments to see the dependence of
    the results on the propagation time (see Section~\ref{sec:ConstancyProperElements}).
\end{itemize}


\subsection{Data analysis of the results}\label{sec:data}
In this Section we implement statistical methods for data analysis
(\cite{stat1}, \cite{stat2}) to quantify the links between the
distributions of osculating, mean and proper elements. As in the
previous Sections, we compare the distributions of semimajor axis,
eccentricity and inclination at break-up, after 150 years, and by
computing the proper elements after 150 years.

Our procedure relies on the following steps: (S1) we use
histograms to check the distributions of the data, (S2) we scan
the datasets to find possible outliers, (S3) we perform the
Kolmogorov-Smirnov (K-S) test to compare the distributions,
and finally (S4) we compute the Pearson correlation coefficient
between the datasets (see Appendix B for further
details on each step of the procedure and the relevant definitions).

As an example, we take the case of moderate orbits presented
in Figure~\ref{fig:ModAt2FullRange} and we implement the above
steps (S1)-(S4) to analyze the data. Since the semi-major axis is
always constant, we are interested just in the analysis of
eccentricity and inclination.

\begin{figure}[h]
\centering
\includegraphics[width=0.8\columnwidth]{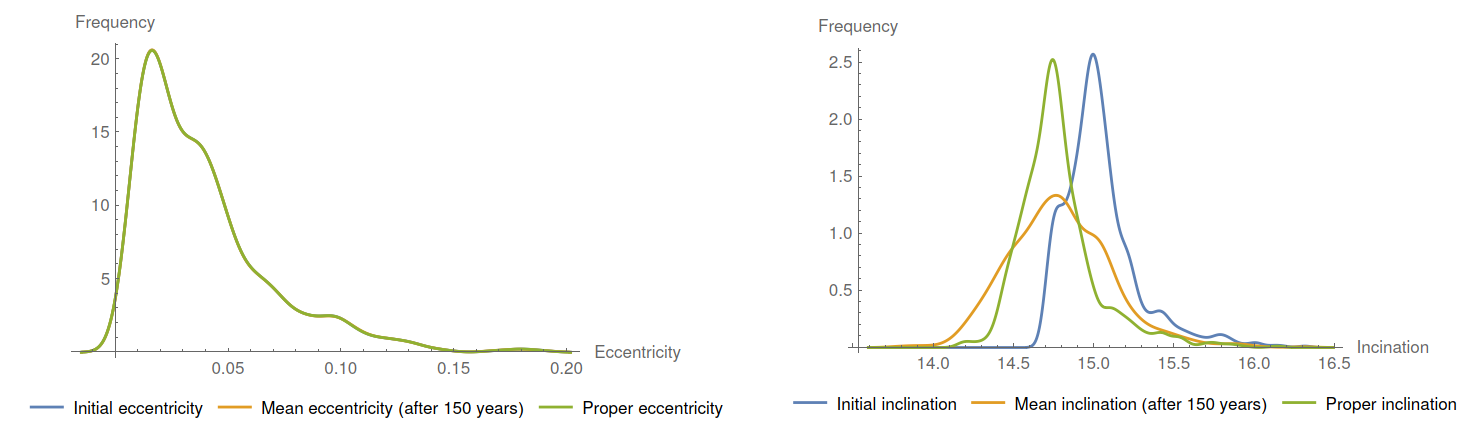}
\caption{Histograms of $e$ (left) and $i$ (right) for a parent
body with orbital elements $a= 20600$ km, $e=0.01$, $i=15^o$,
$\omega = 10^o$, $\Omega = 20^o$.} \label{fig:Hist150yrs}
\end{figure}

In Figure~\ref{fig:Hist150yrs} we show the histograms of $e$
and $i$ (step (S1)) in three situations: initial distribution,
mean elements distribution (after 150 years), and proper elements
distribution (after 150 years). The right plot of Figure~\ref{fig:Hist150yrs} shows the
inclinations: the initial and proper inclinations have the same
shape and size, although they are shifted; the mean inclination
curve has instead different shape and size, thus underlining (once
more) the different behavior of the mean elements with respect to
the initial ones. As for the eccentricity given in the left plot of Figure~\ref{fig:Hist150yrs}, all curves are almost
overlapping, since the forces (geopotential, Sun and Moon) do not
affect too much the evolution of the eccentricity, since we are taking the initial data in a stable region far from the tesseral and lunisolar resonances.

As for step (S2), while checking the outliers for this
experiment, we found 2 anomalies in the dataset of the initial
semi-major axis; however, only one of them is preserved for the
mean and proper semi-major axes. As for the eccentricity, we found
4 outliers in the initial dataset; 3 of them are also found in the
proper eccentricity set, while for the mean eccentricity we found
3 more outliers. Concerning the inclination, we have 3 outliers in
the initial data, 7 in the final mean dataset, and 6 in the proper
inclination set. The conclusion is that the number of outliers is
very small for each dataset and it does not affect the performance
of the other statistical tests (e.g., Pearson correlation and the
K-S test).

The behaviour of the distributions is confirmed also by the
K-S test (step (S3)), which gives a small p-value equal to
$0.0015$ when checking the similarity between the initial
dataset and the mean elements after 150 years, while it gives a
higher p-value equal to $0.5167$ when looking at the initial
data and the proper elements (see Table~\ref{tab:DataAnalysisFull}).

Step (S4) provides evidence of the difference in inclination
between the initial data and the data after 150 years; this is
obtained by computing the Pearson correlation coefficient which
turns out to be equal to $0.7423$. Instead, a higher coefficient
equal to $0.9720$ is obtained when comparing the initial and
proper elements.

We summarize in Table~\ref{tab:DataAnalysisFull} the results of
the data analysis of the different samples studied in the previous
sections; the samples can be identified through the semi-major
axis.

\subsection{The effect of Solar radiation pressure (SRP)}
\label{sec:SRP} Solar radiation pressure affects mostly the
fragments with high area-to-mass ($A/m$) ratio. Since the break-up
simulator returns the $A/m$ ratio for each fragment, we perform
also a test on the computation of the proper elements including SRP.
Following \cite{HughesI}, we use \eqref{eq:SRP} to model SRP; the explicit
expression is given in Appendix C.

Since the Hamiltonian in Appendix C depends also
on the angle $M_S$, we treat the new angle as we did it for the
ascending node of the Moon, namely we introduce a dummy variable
associated to $M_S$ which gives a 5 degrees of freedom Hamiltonian
system. The normalization algorithm then follows the same steps as
described in Section~\ref{sec:NRR}.

\begin{figure}[h]
\centering
\includegraphics[width=0.7\columnwidth]{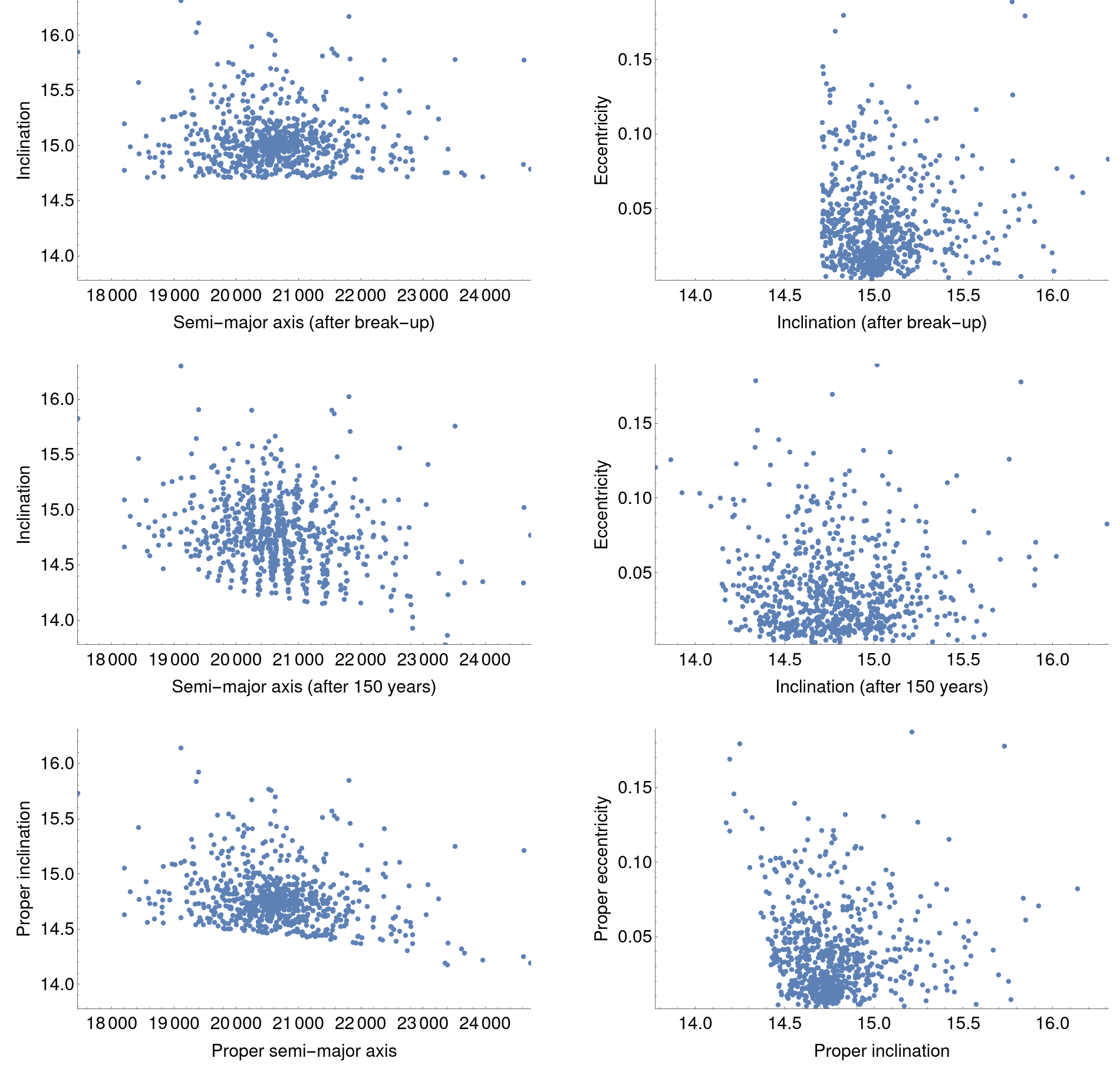}
\caption{Distribution of $a$-$i$ (left), $i$-$e$ (right) for a
parent body with orbital elements $a= 20600$ km, $e=0.01$,
$i=15^o$, $\omega = 10^o$, $\Omega = 20^o$ after break-up (first
row), after 150 years (second row), and proper elements computed
after 150 years (third row). Additional effect: SRP. }
\label{fig:SRPFullRange}
\end{figure}

The results provided in Figure~\ref{fig:SRPFullRange}
reproduce those of Figure~\ref{fig:ModAt2FullRange} with the
addition of SRP, which is computed by the simulator for each
fragment; the values of the area-to-mass ratio range from
$A/m=0.05$ to $A/m=0.74$. Since such values are not too large,
there are not big differences between
Figure~\ref{fig:ModAt2FullRange} and
Figure~\ref{fig:SRPFullRange}. Indeed, we notice only a small
difference in the computation of the Pearson correlation
coefficients between the initial and mean elements ($0.742391$),
and initial and proper elements ($0.932413$). For all other
experiments described in the previous Sections, the values of the
Pearson coefficient and the K-S test p-value are provided in
Table~\ref{tab:DataAnalysisFull}.

\subsection{Proper elements and random noise}
\label{sec:RandomNoises} To complement the previous results, we
add a simple experiment to take into account the behaviour of
proper elements in case of noisy initial conditions. We believe
that this topic deserves a thorough study; however, we think
interesting to give a preliminary insight on a test sample, following this procedure:
consider the evolution of the mean elements
starting from some initial values, add noise to the latter, and
then compute the proper elements using the noisy data.

The noise is introduced by computing the orbital elements
according to the following formulae: $a_{N} = a (1+ 0.001\tau)$,
$e_{N} = e (1+ 0.05\tau)$, $i_{N} = i (1+0.005\tau)$, $M_{N} = M (1+
0.005\tau)$, $\omega_{N} = \omega (1+ 0.005\tau)$, $\Omega_{N} =
\Omega (1+ 0.005\tau)$, where $\tau$ is a random number taking the values -1, 0, or 1.

\begin{figure}[h]
\centering
\includegraphics[width=0.9\columnwidth]{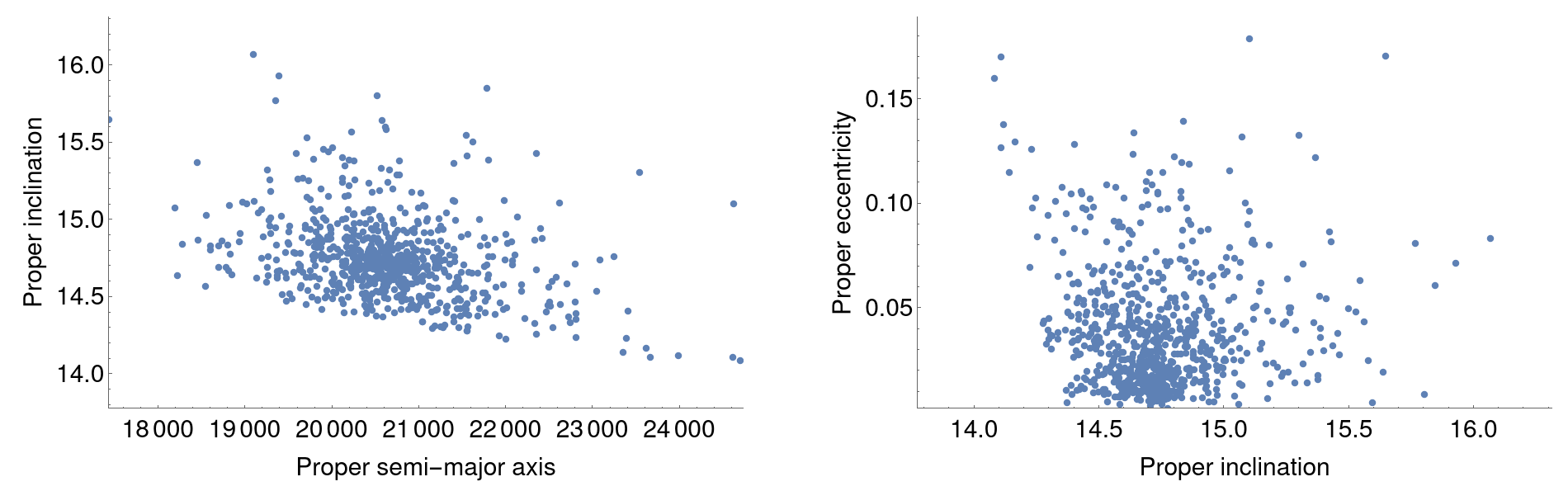}
\caption{Distribution of $a$-$i$ (left), $i$-$e$ (right) with the
parent body at the initial position $a= 20600$ km, $e=0.01$,
$i=15^o$, $\omega = 10^o$, $\Omega = 20^o$: proper elements computed
after 150 years. Additional effect: random noise.}
\label{fig:RandomNoiseFullRange}
\end{figure}

\begin{figure}[h]
\centering
\includegraphics[width=0.7\columnwidth]{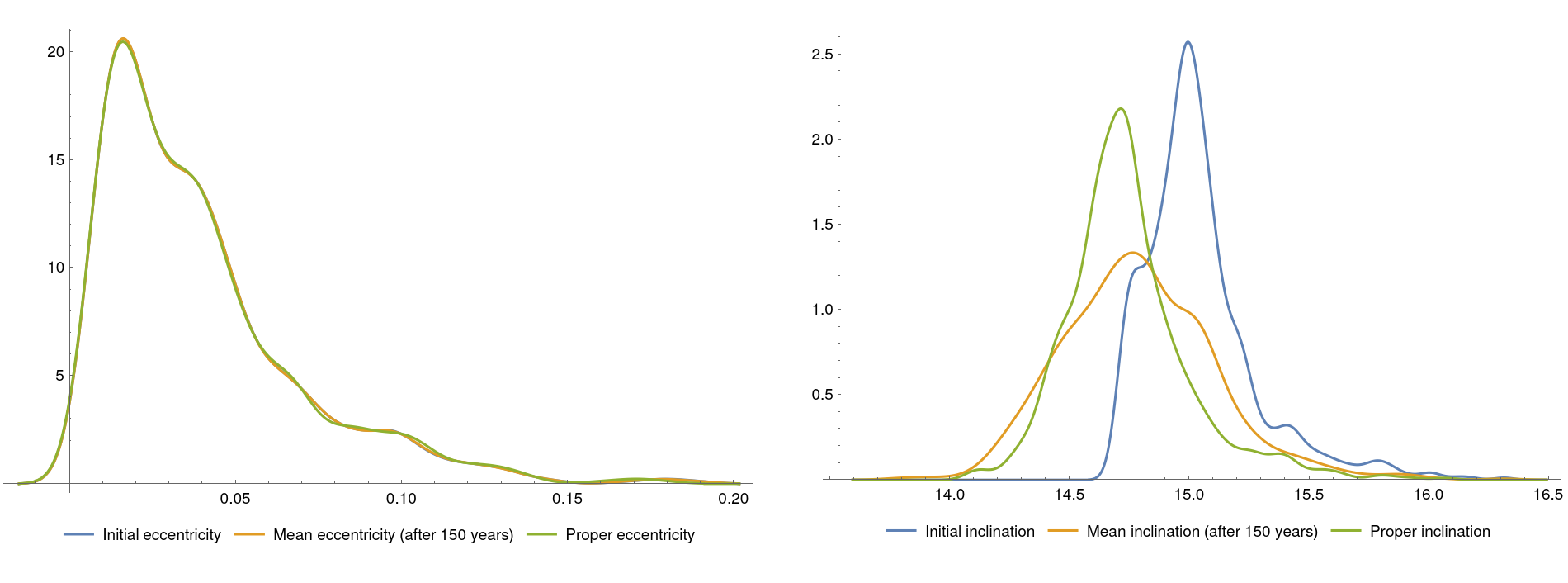}
\caption{Histogram of $e$ (left) and $i$ (right) for e parent body
at the initial position $a= 20600$ km, $e=0.01$, $i=15^o$, $\omega
= 10^o$, $\Omega = 20^o$. Additional effect: random noise.} \label{fig:HistRandomNoise}
\end{figure}

Again, we consider the sample already presented in Figure~\ref{fig:ModAt2FullRange}, but adding noise.
The results show a larger spread with respect to the case without noise as shown in Figure~\ref{fig:RandomNoiseFullRange}, and comparing the histogram from Figure~\ref{fig:Hist150yrs} and Figure~\ref{fig:HistRandomNoise}. The conclusion of such
experiment is that adding a small noise to the orbital elements,
we are still able to reproduce the initial distributions with a
fairly good approximation. A result which is confirmed also by the
values of the statistical tests reported in
Table~\ref{tab:DataAnalysisFull}.

\subsection{Constancy of the proper elements} \label{sec:ConstancyProperElements}
\begin{figure}[h]
\centering
\includegraphics[width=0.9\columnwidth]{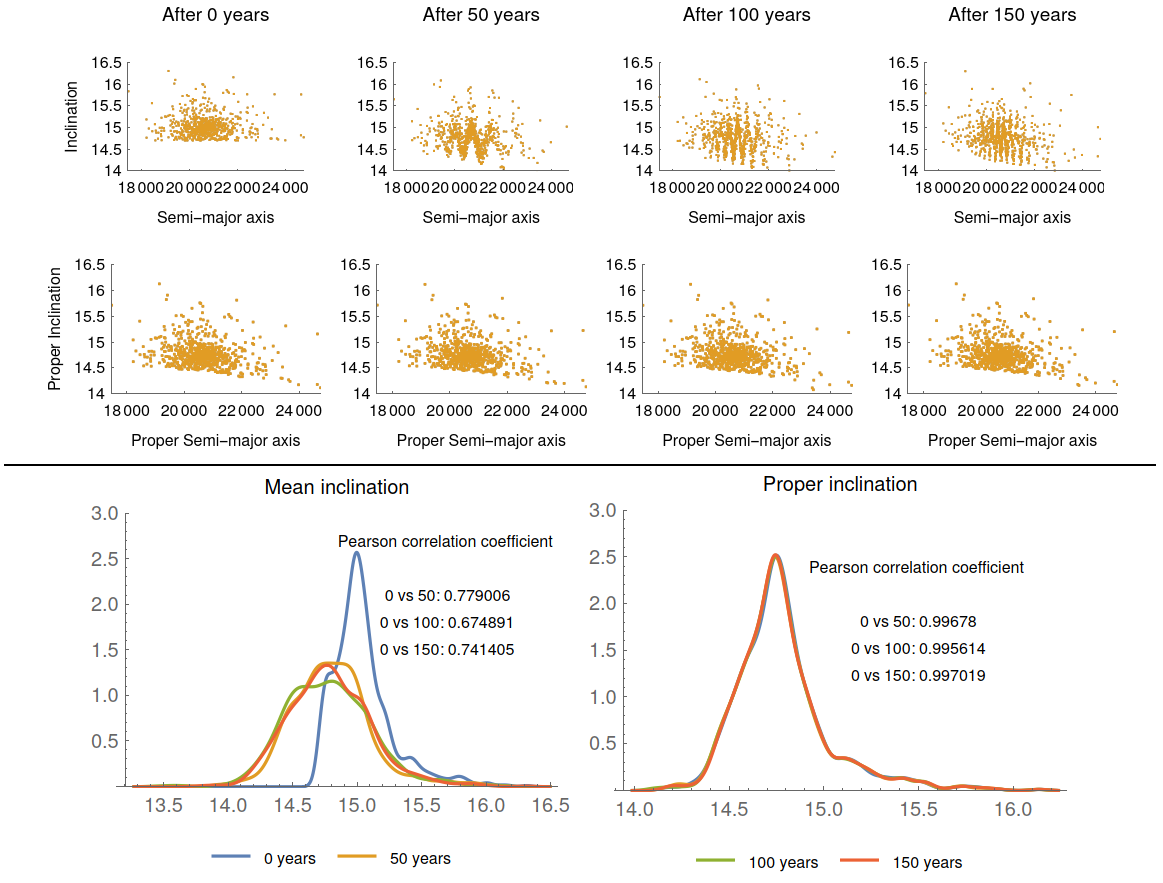}
\caption{Comparison of mean inclination and proper inclination at different times 0
years, 50 years, 100 years, 150 years (top). Distribution and Pearson correlation coefficient for mean
inclination (bottom-left) and for proper inclination (bottom-right).}
\label{fig:Constancy}
\end{figure}

Using the results obtained in Section~\ref{sec:ProperElementsLow},
we analyze how the distribution of the proper elements changes
when we take different evolution times. To this end, we consider the
fragments at four different times: 0 years, 50 years, 100 years
and 150 years. For each time and each fragment, we obtain a set of
final data from which we compute the proper elements; for this
reason, we will refer to the time at which we compute the proper
elements as the \sl proper-time. \rm The results are given in
Figure~\ref{fig:Constancy}. The plots make clear that the mean
elements distributions change with time, while those of the proper
elements are kept nearly constant; this fact is reflected also by
the statistical data analysis shown in Figure~\ref{fig:Constancy},
namely by the histograms and the values of the Pearson correlation
coefficients.

\section{Conclusions and further work}\label{sec:conclusions}
In the present paper we developed a method to compute proper elements for groups of
fragments of space debris associated to the same break-up event. The Hamiltonian formulation of the model
was adopted to describe the dynamics taking into account three main forces:
the potential of the Earth, and the attraction of the Moon and the
Sun. We also make some experiments to evaluate the effect of Solar radiation pressure
and that of data affected by noise.
Using a break-up simulator reproducing the model described in \cite{Johnson2001}, we analyzed the effectiveness of
the computation of the proper elements in different regions, which include moderate altitude
orbits ($a=15100$ km and $a=20600$ km), and
medium altitude orbits ($a=33600$ km). The experiments consisted in a
comparison of the distribution of the elements $a,e,i$ in three
different scenarios: after break-up, after 150 years and
after the transformation into proper elements of the data propagated to 150 years. The main results
come from the plots in Sections~\ref{sec:ProperElementsLow},
\ref{sec:ProperElementsMed}.
The plots obtained computing the proper elements show a
distribution similar to that after the break-up; the results
are supported by data analysis, mainly through histograms, the K-S test and the determination
of the Pearson correlation coefficient. We
have also noticed that the proper elements become very useful
when analyzing the dynamics after long periods of time. Indeed, we can use them as a
prediction of the mean position of the group of fragments using
the information at the break-up time.

We also included a computation of proper elements in the neighborhood of the 1:1 and 2:1
tesseral resonances. Although the present results display already the correct behaviour, we believe that
the resonant cases need a more accurate investigation and their analysis can be improved
through the development of a resonant normalization procedure.

\vskip.1in

Our conclusion is that the present work should be seen as a proof of the conceptual relevance of normal forms and proper elements
in classifying space debris, especially when associated to break-up events, either an explosion or a collision between
two satellites. Some real cases have been successfully studied in \cite{CPV}, using some of the methods detailed in the present work.
In view of possible applications, we foresee several ways to improve
the results, in primis the study of a more elaborated model including a larger number
of spherical harmonics and a higher order expansion of the Hamiltonian. Another
ingredient that can contribute to improve the results is to push the normalization procedure
to higher orders. Besides, we believe that it could be worth investigating the
effectiveness of synthetic versus analytic proper elements. As a further direction
of our work, we plan to explore the construction and use of approximate integrals
within the LEO region, in which the objects are affected by the atmospheric
drag. The dissipative effect in LEO might require to develop tools different
than those presented  in this work.

\footnotesize{
\begin{table}[ht]
\begin{tabular}{|p{2.5cm}|p{1.5cm}|p{3cm}|p{3cm}|p{3cm}|p{3cm}|}
  \hline
  Semi-major axis & Additional effect & K-S p-value inclination after\newline 150 years &  K-S p-value \newline proper inclination &  Pearson coefficient \newline inclination after 150 years&  Pearson coefficient \newline proper inclination \\
  \hline
   & None & $0.000157468$ & $0.504867$ & $0.683247$ & $0.99892$  \\
  $15100$ km & SRP & $0$ & $0$ & $0.397109$ & $0.42841$  \\
   & Noise & $0.000157468$ & $0.00403839$ & $0.683247$ & $0.872222$  \\
  \hline
   & None & $0.00158629$ & $0.516702$ & $0.742373$ & $0.972032$ \\
  $20600$ km & SRP & $0.691085$ & $0.910099$ & $0.742391$ & $0.932413$ \\
   & Noise & $0.00158629$ & $0.223151$ & $0.742373$ & $0.924797$ \\
  \hline
   & None & $0.00543351$ & $0.994865$ & $0.731604$ & $0.999159$ \\
  $24600$ km \newline (2 groups)& SRP & $0.00405436$ & $0.994865$ & $0.738282$ & $0.999185$ \\
   & Noise & $0.00543351$ & $0.994865$ & $0.731604$ & $0.987192$ \\
  \hline
   & None & $0.0137999$ & $0.999994$ & $0.856875$ & $0.999546$ \\
  $24600$ km \newline (3 groups)& SRP & $0.0203053$ & $0.999993$ & $0.862371$ & $0.999562$ \\
   & Noise & $0.0137999$ & $0.968287$ & $0.856875$ & $0.994158$ \\
  \hline
   & None & $0.188238$ & $0.999855$ & $0.126101$ & $0.997886$ \\
  $26600$ km & SRP & $0.159541$ & $0.81169$ & $0.126514$ & $0.786289$ \\
   & Noise & $0.188238$ & $0.910099$ & $0.126101$ & $0.908531$ \\
  \hline
   & None & $0.445528$ & $0.945333$ & $0.22445$ & $0.942989$ \\
  $33600$ km & SRP & $0.445528$ & $0.945333$ & $0.22445$ & $0.942989$ \\
   & Noise & $0.447331$ & $0.447331$ & $0.226055$ & $0.837718$ \\
  \hline
   & None & $0$ & $0.165985$ & $0.0272665$ & $0.36087$ \\
  $40600$ km & SRP & $0$ & $0.0902647$ & $0.0186929$ & $0.349326$ \\
   & Noise & $0$ & $0$ & $0.0680719$ & $0.269618$ \\
  \hline
 \end{tabular}
 \vskip.1in
 \caption{The values for K-S test and Pearson correlation coefficient, obtained by comparing
 mean and proper elements with the initial data.}\label{tab:DataAnalysisFull}
\end{table}
}

\normalsize

\section*{Appendix A - Cartesian equations of motion} \label{Appendix}
Let us consider a body orbiting around the Earth. We assume that it is subject
to the gravitational attraction of the Earth, which we consider as an extended body, the Moon, and the Sun. We consider two reference systems, with origin in the center of the Earth:
\begin{enumerate}
    \item a quasi-inertial frame, where the unit vectors $\{ \underline{e}_1, \underline{e}_2, \underline{e}_3 \}$ are fixed;
    \item a co-rotating frame $\{ \underline{f}_1, \underline{f}_2, \underline{f}_3 \}$, in which the Earth is fixed and $\underline{f}_3$
    is aligned with the spin axis, which rotates at a constant rate $\omega$.
\end{enumerate}
Let $\underline{r}$ be the position of the body, which can be written in both reference systems as
\beqano
\begin{split}
     \underline{r} & = x \underline{e}_1 +
                    y \underline{e}_2 +
                    z \underline{e}_3 \\
                & = X \underline{f}_1 +
                    Y \underline{f}_2 +
                    Z \underline{f}_3\ .
    \end{split}
\eeqano
Denoting by $\theta$ the sidereal time, the relation between the coordinates $(x, y, z)$ and $(X, Y, Z)$ is given by
\beqano
    \begin{pmatrix}
    x \\ y \\ z
    \end{pmatrix}
    = R_3(-\theta)
    \begin{pmatrix}
    X \\ Y \\ Z
    \end{pmatrix},
    \label{eq:rel_vectors}
\eeqano
where
\[
R_3(\theta) =
\begin{pmatrix}
\cos\theta & -\sin\theta & 0 \\
\sin\theta &  \cos\theta & 0 \\
0          &    0        & 1 \\
\end{pmatrix}.
\]
Let $\underline{r}_S, \underline{r}_M$ be the positions of the Sun and the Moon with respect to the center of the Earth,
$m_S$, $m_M$ the masses of the Sun and the Moon, and $\mathcal{G}$ the gravitational constant.
The Cartesian equations of motion for a satellites in orbit around the Earth are written as
\begin{equation}  \label{eq:equationAroundEarth}
\begin{split}
    \ddot{\underline{r}}  =& -\mathcal{G} \int_{{\mathcal{V}_E}} \rho(\underline{r}_p) \frac{\underline{r}-\underline{r}_p}{\lvert
    \underline{r}-\underline{r}_p\rvert^3}d{\mathcal{V}_E}
     - \mathcal{G}m_S \bigg( \frac{\underline{r}-\underline{r}_S}{\lvert\underline{r}-\underline{r}_S\rvert^3} +
     \frac{\underline{r}_S}{\lvert\underline{r}_S\rvert^3}\bigg) \\
    & - \mathcal{G}m_M \bigg( \frac{\underline{r}-\underline{r}_M}{\lvert\underline{r}-\underline{r}_M\rvert^3} +
    \frac{\underline{r}_M}{\lvert\underline{r}_M\rvert^3}\bigg),
\end{split}
\end{equation}
where ${\mathcal{V}_E}$ is the volume of the Earth, $\underline{r}_p$ is the position vector of a point inside the Earth, $\rho(\underline{r}_p)$
denotes the density at $\underline{r}_p$; the other two terms in \equ{eq:equationAroundEarth} represent the gravitational
attraction of Sun and Moon, respectively.
Denoting by
$$
    V(\underline{r}) = \mathcal{G} \int_{\mathcal{V}_E}  \frac{\rho(\underline{r}_p)}{\lvert \underline{r}-\underline{r}_p\rvert}d\mathcal{V}_E
$$
the gravitational potential of the Earth and with $\nabla_{\underline{f}}$ the gradient with respect to the co-rotating frame,
the first term at the right hand side of (\ref{eq:equationAroundEarth}) can be written as
\begin{equation}
R_3(-\theta)  \nabla_{\underline{f}} V(\underline{r}).
\label{eq:rotatingFrame}
\end{equation}

\subsection*{A.1 Spherical harmonics expansion of the geopotential}
The Earth's gravitational potential can be expanded as a series of spherical harmonics
(\cite{kaula1966}). Let us introduce spherical coordinates in the co-rotating frame as
\beqano
    \begin{cases}
    X = r \cos\phi \cos\lambda, \\
    Y = r \cos\phi \sin\lambda, \\
    Z = r \sin\phi, \\
    \end{cases}
\eeqano
where $0 \leq \lambda < 2\pi$, $-\frac{\pi}{2} \leq \phi < \frac{\pi}{2}$.
The series expansion of $V$ in spherical harmonics is given by
$$
    V(r, \phi, \lambda) = \frac{\mathcal{G}m_E}{r}
    \sum_{n=0}^{\infty} \bigg( \frac{R_E}{r} \bigg)^n
    \sum_{m=0}^{n} P_{nm}(\sin\phi)\big[ C_{nm} \cos(m\lambda) + S_{nm}\sin(m\lambda) \big].
$$
Here, the functions $P_{nm}$ are defined in terms of the Legendre polynomials $P_n(x)$ as
$$
    P_{nm}(x) = (1-x^2)^{m/2} \frac{d^m}{dx^m}P_n(x),
$$
and $C_{nm}, S_{nm}$ are the harmonic coefficients obtained by the following formulae:
$$
C_{nm} = \frac{2-\delta_{0m}}{m_E} \frac{(n-m)!}{(n+m)!}\int_{{\mathcal{V}_E}}(\frac{r_p}{R_e})^n P_{nm}(\sin\phi_p)\cos(m\lambda_p)\rho(\underline{r}_p)d{\mathcal{V}_E}
$$
$$
S_{nm} = \frac{2-\delta_{0m}}{m_E} \frac{(n-m)!}{(n+m)!}\int_{{\mathcal{V}_E}}(\frac{r_p}{R_e})^n P_{nm}(\sin\phi_p)\sin(m\lambda_p)\rho(\underline{r}_p)d{\mathcal{V}_E}\ ,
$$
where $m_E$ is the mass of the Earth, $(r_p,\phi_p,\lambda_p)$ denote the spherical coordinates associated to a point $P$ inside the Earth and, again, $r_p$ is its radius vector ($\delta_{jm}$ is the Kronecker symbol).

The coefficients $C_{nm}$ and $S_{nm}$ are related to the quantities $J_{nm}$ and $\lambda_{nm}$ of Table~\ref{table:J} by
\beqano
J_{nm}&=&(C_{nm}^2+S_{nm}^2)^{1\over 2}\qquad {\rm for}\ m\not=0\ , \nonumber\\
J_{n0}&=&J_n=-C_{n0}\quad\ \qquad {\rm for}\ m=0\ .
\eeqano

\subsection*{A.2 Equations of motion up to 2nd order}
The Cartesian equations of motion in the quasi-inertial reference frame are given by the rotation (\ref{eq:rotatingFrame}) and
by computing the partial derivatives of the potential with respect to spherical coordinates:

\begin{equation}
R_3(-\theta)  \nabla_{\underline{f}} V(\underline{r}) = \big(\frac{\partial V}{\partial X} \cos\theta - \frac{\partial V}{\partial Y} \sin\theta\big)
\underline{e}_1 + \big(\frac{\partial V}{\partial X} \sin\theta + \frac{\partial V}{\partial Y} \cos\theta\big)
\underline{e}_2+\frac{\partial V}{\partial Z}\underline{e}_3\ ,
\label{eq:potentialRotatingFrame}
\end{equation}
where the potential up to order $n=m=2$ in the co-rotating frame is given by
$$
V(X,Y,Z) = \frac{\mathcal{G} m_E}{r}+\frac{\mathcal{G} m_E}{r}(\frac{R_E}{r})^2\big[C_{20}\big(\frac{3Z^2}{2r^2}-\frac{1}{2}\big)+3C_{22}\frac{X^2-Y^2}{r^2}+6S_{22}\frac{XY}{r^2}\big]\ .
$$

By differentiating the potential $V$ with respect to $X,Y,Z$ and substituting
the result in (\ref{eq:potentialRotatingFrame}) and (\ref{eq:equationAroundEarth}),
we find the following expression for the Cartesian equations of motion:
\begin{equation*}
\begin{split}
\ddot{x} = &-\frac{\mathcal{G} m_E x}{r^3} + \frac{\mathcal{G} m_E R_E^2}{r^5}  \left\{ C_{20} \big( \frac{3}{2} x - \frac{15}{2} \frac{xz^2}{r^2} \big) +6C_S^{-} x+6C_S^{+} y \right. \\
& \left. + \frac{15x}{r^2} [ C_S^{-} (y^2-x^2)-2xyC_S^{+} ] \right\} - \mathcal{G}m_S \bigg( \frac{x-x_S}{\lvert \underline{r}-\underline{r}_S\rvert^3} +
\frac{x_S}{\lvert x_S\rvert^3}\bigg) \\
&  - \mathcal{G}m_M \bigg( \frac{x-x_M}{\lvert \underline{r}-\underline{r}_M\rvert^3} + \frac{x_M}{\lvert \underline{r}_M\rvert^3}\bigg),
\end{split}
\end{equation*}

\begin{equation*}
\begin{split}
\ddot{y} = &-\frac{\mathcal{G} m_E y}{r^3} + \frac{\mathcal{G} m_E R_E^2}{r^5}  \left\{ C_{20} \big( \frac{3}{2} y - \frac{15}{2} \frac{yz^2}{r^2} \big) +6C_S^{+} x - 6C_S^{-} y \right. \\
& \left. + \frac{15y}{r^2} [ C_S^{-} (y^2-x^2)-2xyC_S^{+} ] \right\} - \mathcal{G}m_S \bigg( \frac{y-y_S}{\lvert \underline{r}-\underline{r}_S\rvert^3} +
\frac{y_S}{\lvert \underline{r}_S\rvert^3}\bigg) \\
&  - \mathcal{G}m_M \bigg( \frac{y-y_M}{\lvert \underline{r}-\underline{r}_M\rvert^3} + \frac{y_M}{\lvert \underline{r}_M\rvert^3}\bigg),
\end{split}
\end{equation*}

\begin{equation*}
\begin{split}
\ddot{z} = &-\frac{\mathcal{G} m_E z}{r^3} + \frac{\mathcal{G} m_E R_E^2}{r^5}  \left\{ C_{20} \big( \frac{9}{2} z - \frac{15}{2}
\frac{z^3}{r^2} \big) + \frac{15z}{r^2} [ C_S^{-} (y^2-x^2)-2xyC_S^{+} ] \right\} \\
& - \mathcal{G}m_S \bigg( \frac{z-z_S}{\lvert \underline{r}-\underline{r}_S\rvert^3} + \frac{z_S}{\lvert \underline{r}_S\rvert^3}\bigg)
- \mathcal{G}m_M \bigg( \frac{z-z_M}{\lvert \underline{r}-\underline{r}_M\rvert^3} + \frac{z_M}{\lvert \underline{r}_M\rvert^3}\bigg)\ ,
\end{split}
\end{equation*}
where $C_S^+=C_{22}\sin 2\theta+S_{22}\cos 2\theta$, $C_S^-=C_{22}\cos 2\theta-S_{22}\sin 2\theta$.

\section*{Appendix B - Data Analysis}\label{app:data}
In this Section we report the basic elements of data analysis,
used in the rest of this work to get useful information from the
dataset associated to the fragments. We refer to \cite{stat1},
\cite{stat2} for further details.

\subsection*{B.1 Histogram}
To visualize the data and to understand the main features of a
distribution, one can plot the histogram of the dataset. This plot
shows the frequency of each element from the set. It turns out to
be a useful tool to compare the distributions of two or more data
sets.

\subsection*{B.2 Outliers}
Outliers are rare data points far away from regular data points and generally do not form a tight cluster.
To check them, we count the anomalies in each dataset of semi-major axis, eccentricity, and inclination in all three
situations: initial osculating elements, final mean elements, and
proper elements. We used a predefined
\texttt{Mathematica}$^\copyright$ function, $FindAnomalies$
from the $MachineLearning$ package, which is computed as:
$$
p = \mathbb{P}(f_{\mathcal{D}}(y) \leq f_{\mathcal{D}}(x), y
\approx \mathcal{D})\ ,
$$
where $\mathbb{P}$ denotes the probability, $\mathcal{D}$ is the
approximated distribution of a 1D array $X$, $x$ is an element in
$X$ and $f_{\mathcal{D}}(x)$ is the probability density function
of $\mathcal{D}$; $y \approx \mathcal{D}$ means that $y$ follows
the distribution $\mathcal{D}$. In case of $p < 0.001$ (default
threshold), $FindAnomalies$ considers $x$ as anomalous. This
function returns the number of outliers, the value of each
outlier, and its position in the dataset. With this information,
we can compare the outliers for osculating, mean and proper
elements.
\subsection*{B.3 Pearson correlation coefficient}
Pearson correlation coefficient, usually denoted by $r$, is
used as a statistical measurement of the relationship between two
one-dimensional datasets. Mathematically, it is a real number in
$[-1,1]$, where $1$ means a total positive linear relationship,
$0$ means no relationship, and $-1$ means a total negative linear
relationship between the two datasets.

For two variables (1D arrays) $X=(x_1,\dots,x_n),
Y=(y_1,\dots,y_n)$, we define the following statistical measures:
\begin{enumerate}
\item Mean of $X$: $\bar{X} = \frac{1}{n}\sum_{i=1}^n x_i$
\item Variance of $X$: $Var[X] = \frac{1}{n-1}\sum_{i=1}^n (x_i - \bar{X})$
\item Covariance of $X$ and $Y$: $Cov[X,Y] = \frac{1}{n-1}\sum_{i=1}^n (x_i - \bar{X})(y_i
-\bar{Y})$\ .
\end{enumerate}
Then, the Pearson correlation coefficient is defined as
$$
r = \frac{Cov[X,Y]}{Var[X]Var[Y]}\ .
$$

\subsection*{B.4 Kolmogorv-Smirnov test}
Kolmogorov-Smirnov test is a goodness-of-fit test where the
null hypothesis says that two datasets are drawn from the same
distribution and the alternative hypothesis that they were not
drawn from the same distribution.

\section*{Appendix C - Solar radiation pressure}\label{app:SRP}
Here we give the full expression of the Solar radiation
pressure Hamiltonian, starting from the expression \eqref{eq:SRP}:
{\footnotesize{
\begin{multline}\label{eq:HamSRP}
\mathcal{H}_{SRP}  =
a \frac{A}{m} \left(\left(-1.1446\times 10^{-10}\right) e \cos \left(-\omega -\Omega -M_S+4.9382\right) \right. \\ \left.
-\left(1.1446\times 10^{-10}\right) e \cos (i) \cos \left(-\omega -\Omega -M_S+4.9382\right)-\left(1.1446\times 10^{-10}\right) e \cos \left(\omega -\Omega -M_S+4.9382\right) \right. \\ \left.
+\left(1.1446\times 10^{-10}\right) e \cos (i) \cos \left(\omega -\Omega -M_S+4.9382\right)-\left(3.2784\times 10^{-6}\right) e \cos \left(-\omega -\Omega +M_S+4.9382\right) \right. \\ \left.
-\left(3.2784\times 10^{-6}\right) e \cos (i)\cos \left(-\omega -\Omega +M_S+4.9382\right)-\left(3.2784\times 10^{-6}\right) e \cos \left(\omega -\Omega +M_S+4.9382\right) \right. \\ \left.
+\left(3.2784\times 10^{-6}\right) e \cos (i) \cos \left(\omega -\Omega +M_S+4.9382\right)-\left(1.4109\times 10^{-7}\right) e \cos \left(-\omega +\Omega +M_S+4.9382\right) \right. \\ \left.
+\left(1.4109\times 10^{-7}\right) e \cos (i) \cos \left(-\omega +\Omega +M_S+4.9382\right)-\left(1.4109\times 10^{-7}\right) e \cos \left(\omega +\Omega
   +M_S+4.9382\right) \right. \\ \left.
-\left(1.4109\times 10^{-7}\right) e \cos (i) \cos \left(\omega +\Omega +M_S+4.9382\right)-\left(1.0959\times 10^{-7}\right) e \cos \left(-\omega -\Omega +2 M_S+4.9382\right) \right. \\ \left. -\left(1.0959\times 10^{-7}\right) e \cos (i) \cos
   \left(-\omega -\Omega +2 M_S+4.9382\right)-\left(1.0959\times 10^{-7}\right) e \cos \left(\omega -\Omega +2 M_S+4.9382\right) \right. \\ \left. +\left(1.0959\times 10^{-7}\right) e \cos (i) \cos \left(\omega -\Omega +2 M_S+4.9382\right)-\left(4.7161\times
   10^{-9}\right) e \cos \left(-\omega +\Omega +2 M_S+4.9382\right) \right. \\ \left. +\left(4.7161\times 10^{-9}\right) e \cos (i) \cos \left(-\omega +\Omega +2 M_S+4.9382\right)-\left(4.7161\times 10^{-9}\right) e \cos \left(\omega +\Omega +2
   M_S+4.9382\right) \right. \\ \left. -\left(4.7161\times 10^{-9}\right) e \cos (i) \cos \left(\omega +\Omega +2 M_S+4.9382\right)-\left(3.0904\times 10^{-9}\right) e \cos \left(-\omega -\Omega +3 M_S+4.9382\right) \right. \\ \left. -\left(3.0904\times 10^{-9}\right) e \cos (i) \cos
   \left(-\omega -\Omega +3 M_S+4.9382\right)-\left(3.0904\times 10^{-9}\right) e \cos \left(\omega -\Omega +3 M_S+4.9382\right) \right. \\ \left. +\left(3.0904\times 10^{-9}\right) e \cos (i) \cos \left(\omega -\Omega +3 M_S+4.9382\right)-\left(1.3299\times
   10^{-10}\right) e \cos \left(-\omega +\Omega +3 M_S+4.9382\right) \right. \\ \left. +\left(1.3299\times 10^{-10}\right) e \cos (i) \cos \left(-\omega +\Omega +3 M_S+4.9382\right)-\left(1.3299\times 10^{-10}\right) e \cos \left(\omega +\Omega +3
   M_S+4.9382\right) \right. \\ \left. -\left(1.3299\times 10^{-10}\right) e \cos (i) \cos \left(\omega +\Omega +3 M_S+4.9382\right)-\left(1.3602\times 10^{-6}\right) e \cos \left(-\omega +M_S+4.9382\right) \sin (i) \right. \\ \left. +\left(1.3602\times 10^{-6}\right) e \cos
   \left(\omega +M_S+4.9382\right) \sin (i)-\left(4.5468\times 10^{-8}\right) e \cos \left(-\omega +2 M_S+4.9382\right) \sin (i) \right. \\ \left. +\left(4.5468\times 10^{-8}\right) e \cos \left(\omega +2 M_S+4.9382\right) \sin (i)-\left(1.2822\times 10^{-9}\right)
   e \cos \left(-\omega +3 M_S+4.9382\right) \sin (i) \right. \\ \left.
   +\left(1.2822\times 10^{-9}\right) e \cos \left(\omega +3 M_S+4.9382\right) \sin (i)\right).\nonumber\\
\end{multline}
} }

\vspace{0.5em}

{\bf Acknowledgements.}
The authors thank C. Efthymiopoulos, C. Gale\c s and D. Marinucci for useful discussions and
suggestions. The authors thank the anonymous Reviewers that greatly contributed to improve the content and the presentation of this work.
All authors acknowledge EU H2020 MSCA ETN Stardust-R Grant Agreement 813644.
A.C. (partially) acknowledges the MIUR Excellence Department Project
awarded to the Department of Mathematics, University of Rome Tor
Vergata, CUP E83C18000100006.
A.C. (partially) and G.P. acknowledge MIUR-PRIN 20178CJA2B
``New Frontiers of Celestial Mechanics: theory and Applications''.
G.P. acknowledges GNFM/INdAM and is partially supported by INFN (Sezione di Roma II).

\vspace{1em}
{\bf Conflict of Interest.} The author A.C. is Editor-in-Chief of the journal ``Celestial Mechanics and Dynamical Astronomy"; the paper underwent a standard single-blind peer review process.

\vspace{1em}
{\bf Data availability.} The datasets generated during and/or analysed during the current study are available from the corresponding author on reasonable request.

\end{document}